%% file: HPCCTechnicalReport.tex
\documentclass[12pt] {article}
\pdfoutput=1
\usepackage{epsfig}
\usepackage{amssymb}
\usepackage{amsmath}
\usepackage{multirow}
\usepackage{setspace}
\usepackage{datetime}

\usepackage{acronym}
\usepackage{array}
\usepackage{graphicx}
\usepackage{cite}
\usepackage{graphicx}
\usepackage{epstopdf}
\usepackage{caption}
\usepackage{subcaption}
\usepackage{balance}
\usepackage{fancybox}
\usepackage{colortbl}
\usepackage{array}
\usepackage{mystyle}
\usepackage{subcaption}
\usepackage{multirow}
\usepackage[obeyspaces,hyphens,spaces]{url}
\usepackage{threeparttable}

\renewcommand{\baselinestretch}{1}

\providecommand{\keywords}[1]{\textbf{\textit{Index terms---}} #1}

\newcommand{\RQone}{How often does a scheduled task or job is Failed, Evicted, Lost, or Killed?}
\newcommand{\RQtwo} {Can we predict the outcome of scheduling events based on cluster log files?}
\newcommand{\RQthree} {Which benefits can be achieved by predicting the outcome of scheduling events?}

\newenvironment{myindentpar}[1]%
{\begin{list}{}%
         {\setlength{\leftmargin}{#1}}%
         \item[]%
}
{\end{list}}

\newdateformat{monthyeardate}{%
  \monthname[\THEMONTH], \THEYEAR}
\begin{document}
\input{abstractnew}

\input{introduction}

\input{background}
\input{casestudy}

\input{model}
\input{results}

\input{ApplicationHadoop}
\input{threats}

\input{relatedwork}

\input{conclusion}
\balance
\bibliography{mabiblio}{}
\bibliographystyle{IEEEtran}
\end{document}

%% file: abstractnew.tex
\title{\textbf{\Large{Predicting Scheduling Failures in the Cloud}}}

\author{\begin{tabular}[t]{c@{\extracolsep{10em}}c}
\multicolumn{2}{c}{\hspace*{1em}Mbarka~Soualhia$^{1}$, Foutse~Khomh$^{2}$ and Sofi\`ene~Tahar$^{1}$} \vspace*{2em}\\
$^{1}$Department of Electrical and Computer Engineering,\\
Concordia University, Montr\'eal, Canada  \\
\{soualhia,tahar\}@ece.concordia.ca \vspace*{2em}\\
$^{2}$Department of Software Engineering,\\
\'Ecole Polytechnique, Montr\'eal, Canada \\
foutse.khomh@polymtl.ca \vspace*{3em}\\
\textbf{TECHNICAL REPORT}\\
\monthyeardate\today
\date{}
\end{tabular}}
\maketitle

\newpage
\begin{abstract}
Cloud Computing has emerged as a key technology to deliver and manage computing, platform, and software services over the Internet. Task scheduling algorithms play an important role in the efficiency of cloud computing services as they aim to reduce the turnaround time of tasks and improve resource utilization. Several task scheduling algorithms have been proposed in the literature for cloud computing systems, the majority relying on the computational complexity of tasks and the distribution of resources. %and the architecture of the systems to derived scheduling decisions.
%However, given the highly dynamic nature of cloud environments,
However, several tasks scheduled following these algorithms still fail because of unforeseen changes in the cloud environments. In this paper, using tasks execution and resource utilization data extracted from the execution traces of real world applications at Google, we explore the possibility of predicting the scheduling outcome of a task using statistical models. %and historical information about the execution of previously scheduled tasks.
If we can successfully predict tasks failures, we may be able to reduce the execution time of jobs by rescheduling failed tasks earlier (\ie{} before their actual failing time). Our results show that statistical models can predict task failures with a precision up to 97.4\%, and a recall up to 96.2\%. We simulate the potential benefits of such predictions using the tool kit GloudSim and found that they can improve the number of finished tasks by up to 40\%. We also perform a case study using the Hadoop framework of Amazon Elastic MapReduce (EMR) and the jobs of a gene expression correlations analysis study from breast cancer research. We find that when extending the scheduler of Hadoop with our predictive models, the percentage of failed jobs can be reduced by up to 45\%, with an overhead of less than 5 minutes. %over a total execution time of 25 minutes. %WeResults show  as an application deployed on Hadoop EC2.}
%implemented our task failures prediction algorithm in  \textcolor{red}{and increase the number of accurate job results by ....}.

\end{abstract}

\keywords{Failure Prediction, Tasks Scheduling, Cloud, Google Clusters, Hadoop, Amazon Elastic MapReduce.}
\newpage
\tableofcontents
%\end{keywords}

%% file: introduction.tex
\newpage
\section{Introduction}
\vspace{-0.1cm}
Cloud Computing has emerged as a key technology that delivers and manages services over the Internet. Customers can lease services provided by cloud computing systems, ramping up or down the capacity as they need and paying only for what they use. Nowadays, cloud computing services are used for several applications such as Internet of Things, Image Processing, Data Mining, and Web Analytics~\cite{[2]}\cite{[1]}.
Task scheduling problems are of paramount importance in cloud environments. Indeed, an efficient scheduling of tasks and jobs across the various heterogeneous virtual clusters that constitute a cloud is critical to ensure good computation time and resource utilisation. Although several task scheduling algorithms have been proposed in the literature for cloud computing systems, cloud schedulers still experience many failures due to unforeseen events such as unpredicted demands of services or hardware outages.%;
We believe that an efficient scheduling of tasks requires a proactive response to changes in cloud environments. If we can predict changes in cloud environments accurately, we may be able to adjust scheduling decisions accordingly and reduce the amount of task scheduling failures. Recently, Chen et al \cite{[21],[22]} examined tasks failures in compute clouds and suggest that predicted failed tasks be killed immediately without processing them, in order to avoid wasting resources. %Their approach is based on a systematic killing of predicted failed jobs \Foutse{tasks? we need to clarify our terminology and make it consistent}.
However, although killing these predicted failed tasks may reduce resources wastage, it does not guarantee a good level of QoS (Quality of Service), since the killed tasks are likely to affect the overall performance of a cloud application. A better decision would be to reschedule the tasks quickly on appropriate clusters with adequate resources in order to ensure their timely and successful completion.

In this paper, we explore the possibility of predicting the scheduling outcome of a task using statistical models and historical information about the execution of previously scheduled tasks. Our goal is to achieve early rescheduling of potential failed tasks in order to improve tasks and jobs execution time and resources utilisation.
We use statistical modelling to establish and inspect dependencies between tasks and jobs characteristics such as execution time, scheduling time, resources usage, machines workload, scheduling constraints, and tasks scheduling outcomes. Using tasks execution and resource utilization data from Google applications, collected over a period of one month in 2011~\cite{[8]}, we address the following three research questions:\\

\vspace{-0.1cm}
\begin{itemize}
\item \textbf{\textit{\RQone }}
\end{itemize}

\begin{myindentpar}{1cm}
We observed that 42\% of the jobs and 40\% of the tasks from the Google dataset were not finished successfully. We also found that unfinished (\ie{} evicted, failed or killed) jobs and tasks are characterized by long waiting times and execution times. Moreover, we noticed that a job often fails because of the failures of some of its tasks, and tasks also fail because of the failure of dependent tasks.
\end{myindentpar}

\begin{itemize}
\item \textbf{\textit{\RQtwo }}
\end{itemize}
\begin{myindentpar}{1cm}
First, we determined the variables that affect directly the scheduling outcome of task or job. Then, we applied Decision Tree, Boost, GLM, CTree, Random Forest and Neural Network algorithms to predict whether or not a scheduled task will fail. Our best prediction model is obtained with Random Forest. This model achieves a precision up to 97.4\%, and a recall up to 96.2\%. Cloud service providers could make use of such prediction models to improve the performance of their scheduling algorithms.

\end{myindentpar}

\begin{itemize}
\item \textbf{\textit{\RQthree }}
\end{itemize}
\begin{myindentpar}{1cm}
We evaluate the potential benefits of our prediction models using the tool kit GloudSim which was built to simulate the original workload of Google applications~\cite{[20]}.
We examine whether our models can identify and predict failure events when scheduling tasks and enable better scheduling decisions.
Results show that prediction models can help reduce the execution time of the jobs and tasks. Also, the early failure predictions reduce the number of failed tasks by up to 40\%.
\end{myindentpar}

To demonstrate the practicality of our prediction models in a real world setting, we implement and deploy the obtained prediction models on Amazon EC2, extending the scheduler of the Hadoop framework of Amazon EMR ~\cite{[20]}. We reproduce and execute a series of jobs from a gene expression correlations analysis study in breast cancer research ~\cite{[Chang-BreastCancer]}. Results show that the extended version of Hadoop's scheduler generates better scheduling policies, \ie{} the percentage of failed jobs is reduced by 45\%. This improvement of the performance of the scheduler is achieved at a minimum cost of less than 5 minutes over a total execution time of 30 minutes.

The remainder of this paper is organized as follows: Section \ref{sec:bachground} gives a general overview about tasks and jobs scheduling. Section \ref{methdology} describes the case study design, our proposed methodology to process the data from Google Traces. Section \ref{sec:casestudy:results} describes the results of \textbf{RQ1} and \textbf{RQ2}. The simulation of the benefits of our proposed prediction models with GloudSim (\ie{} \textbf{RQ3}) is presented in Section \ref{sec:predictionresults}. Section \ref{sec:application} presents the results of the case study with Hadoop. Section \ref{sec:threats} discusses threats to the validity of our work, Section \ref{sec:relatedwork} summarizes the related literature and Section \ref{sec:conclusion} presents the conclusion and discusses future works. %Finally, the conclusion is will be given in .

%% file: background.tex
\section{Background on Jobs and Tasks Scheduling}
\label{sec:bachground}
\begin{figure}[th]
\centering
\includegraphics[scale=.35]{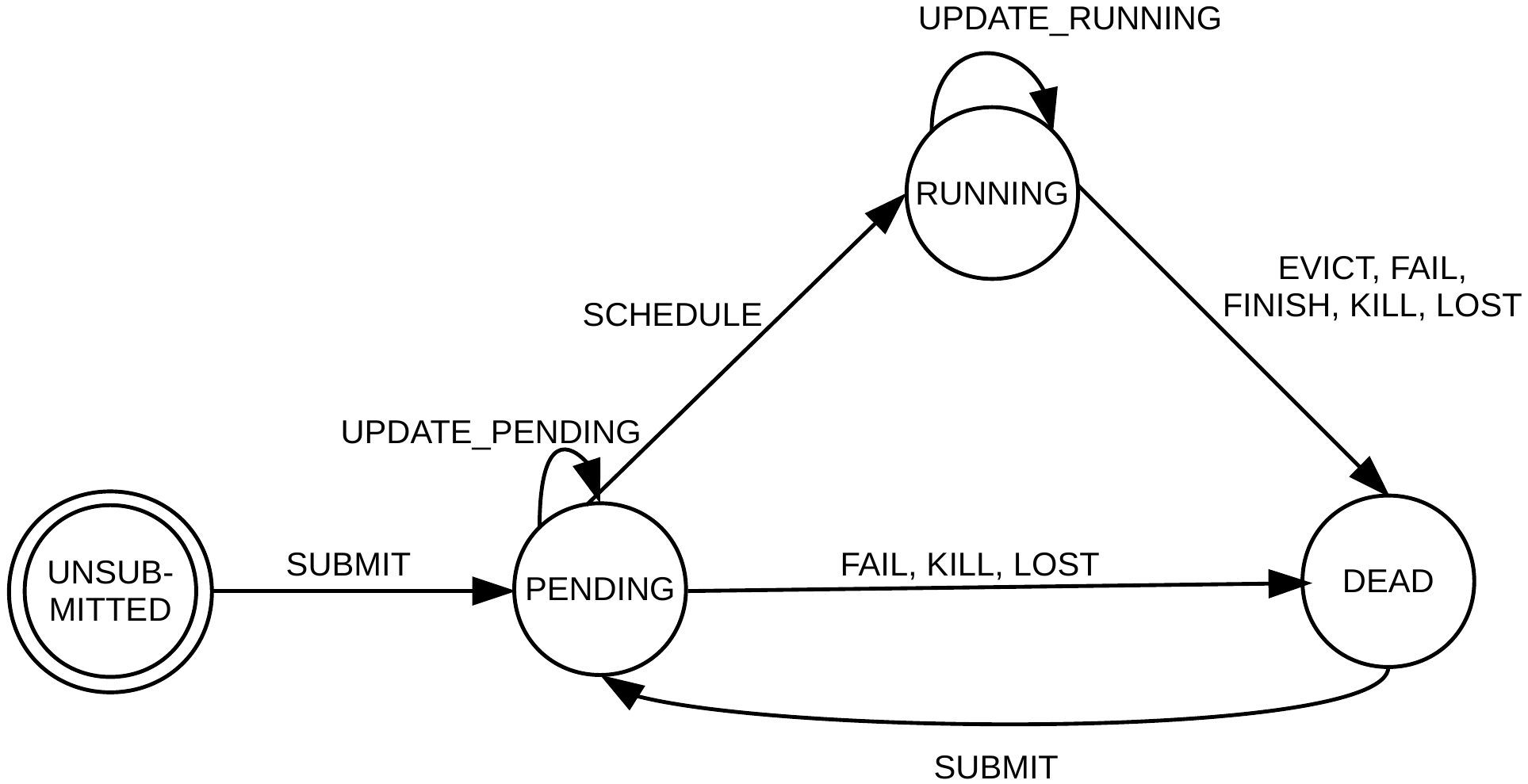}
\caption{State Transitions of Google Tasks and Jobs \cite{[8]}}
\label{fig:statetransition}
\vspace{-15pt}
\end{figure}

In a cloud environment composed of multiple clusters, task scheduling aims to allocate a number of dependent and/or independent tasks with a given execution time to the machines having enough resources in the clusters. A good scheduler can minimize the execution time and improve the utilisation of the allocated resources~\cite{[24]}. In particular, the main goal of a scheduler is to find the optimal solution to schedule the submitted tasks to the proper virtual machines in accordance with the optimal execution time and resources availability. So, the scheduler will look for the machine or the processor having the minimum of the required resources to process the tasks to satisfy their requirements while reducing the resources utilisation cost.
%\Foutse{while reducing the resources utilisation cost}. 
Task scheduling is one of the most important problems when implementing real time applications~\cite{[17]}. %and many research algorithms have been proposed to solve it \cite{[17]}.
%However, this hot issue is complex and known as a
It is a combinatorial optimization problem since it involves multiple complex variables and constraints on a large scale. %in cloud computing environment.
According to the description of Google traces in which users' applications are considered as jobs composed of one or more tasks, the typical scheduling life cycle of a task or job is composed of four different states as shown in Figure~\ref{fig:statetransition}.
%
%Many techniques and algorithms have been proposed to enhance schedulers efficiency by studying . \\
%In this context, we will describe as an example the scheduling life cycle used in Google distributed clusters \cite{[8]} in which users applications can be considered as jobs composed of one or more tasks. Tasks and jobs have scheduling life cycle composed of four different states as shown in Figure~\ref{fig:statetransition}. According to Google traces description,
Each task can only be in one of the following states: \textit{Unsubmitted, Pending, Running or Dead}. The transition between two states occurs on the scheduler or the processing machines %on the machine where the job or the task is processed, 
only when a task scheduling event occurs. There are 9 types of task scheduling events : \textit{Submit, Schedule, Evict, Fail, Finish, Kill, Lost, Update$-$Pending and Update$-$Running}. Any submitted task will be waiting in the queue for some available resources that meet its constraints and then will be scheduled and assigned to the appropriate processor for execution. A task or job can be Failed or Killed before its completion. %being successfully finished.
Tasks that are killed before their completion or that failed to be submitted are resubmitted again. 
%\Foutse{how many times? --> it depends on the implementation of cluster scheduler: this info is not described in the description of the data}. 
The Google scheduler uses the priority of tasks or jobs to make scheduling decisions. In case of multiple tasks or jobs having the same priority, scheduling decisions are made based on the order of arrival of the tasks. %using the FCFS scheduling principles to process jobs ans tasks by the order of their arrival.
The first task to arrived is served first and the next one is queued until there are available resources in the cluster. More details about task scheduling in Google clusters can be found in \cite{[8]}.

%% file: casestudy.tex
\section{Methodology}
\label{methdology}
\begin{figure*}[th]
\centering
\includegraphics[scale=.4]{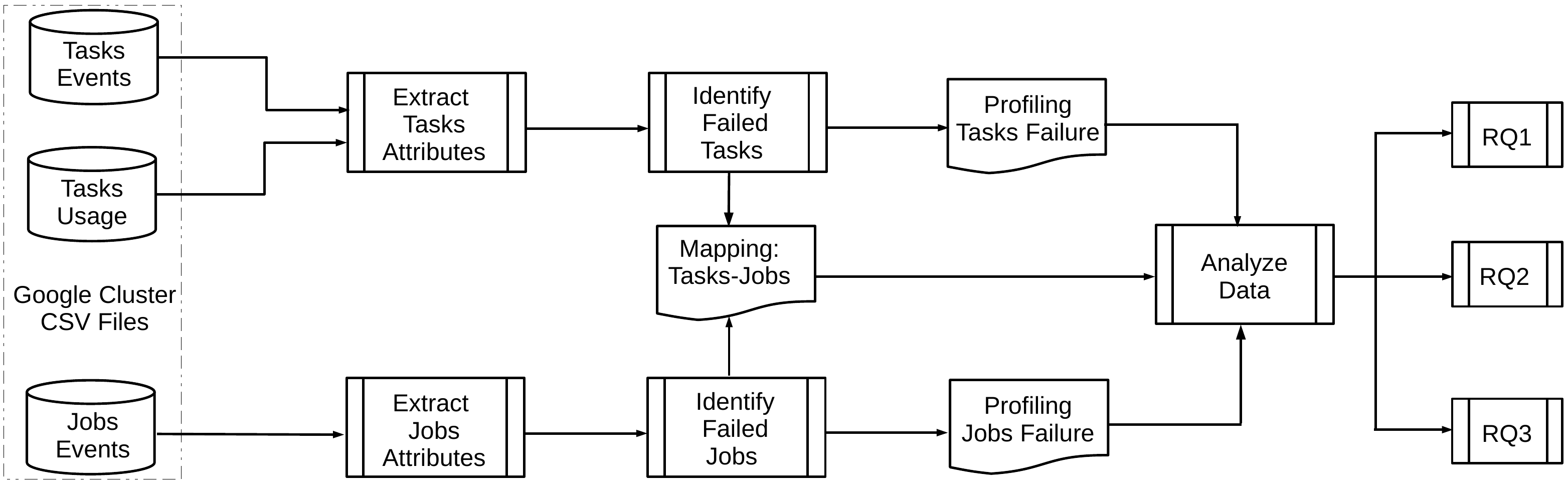}
\caption{Overview of Our Approach to Extract Data from Google Cluster CSV Files}
\label{Figure:overview}
\vspace{-5pt}
\end{figure*}
In this section, we will describe the design of our study, the studied system and our data extraction and analysis approaches to answer the following research questions:
\begin{itemize}
\item \textit{RQ1)} \RQone
\end{itemize}
\begin{itemize}
\item \textit{RQ2)} \RQtwo
\end{itemize}
\begin{itemize}
\item \textit{RQ3)} \RQthree
\end{itemize}

\subsection{Case Study Design}
In order to answer our three research questions, we performed an empirical study using large-scale data (i.e., \textit{158 GB}) collected from Google clusters. %\Foutse{can you revise the following answer?--> I reworded their definition about their cluster}
A Google cluster is a set of different machines that are inter-connected with high-bandwidth network dedicated to large and distributed clusters. The machines of one cluster are sharing the same scheduling resources allocation and management systems. The schedulers of these machines receive and schedule a large number of jobs (i.e., users' applications). A job is composed of one or multiple tasks~ \cite{[8]}. Jobs are classified into four categories : \textit{single$-$task, sequential$-$tasks, parallel$-$tasks and mix$-$mode$-$tasks}. Each task is a Linux program involving one or %\Foutse{do you mean the jobs in your dataset? can you explain?: No this is how a task and job are described on the document of google}
multiple processes. Every task or job has its own resources requirements in terms of \textit{CPU}, \textit{RAM} and \textit{Disk Space}, and its own scheduling priority and constraints. \\
The Google dataset contains six tables in CSV (Comma-Separated Values) format: \textit{Task Event, Job Event, Machine Attribute, Task Constraint, and Task Usage}. Table~\ref{Table:Google:Traces:Structure} summarises statistics about the content of these tables. %including scheduling events and constraints, resources usage, and machine attributes and events produced and collected during one month in Google clusters.
The information contained in the tables are not fully complete; in fact Google reported that some information about inter-job dependencies, resources types, resource usages and constraints were omitted because of security and confidentiality issues~\cite{[8]}.

\begin{table*}[ht]
\caption{Google Traces Structure and Content}
\centering \scriptsize
\label{Table:Google:Traces:Structure}
\begin{tabular}{|>{\centering\arraybackslash}p{2.5cm}|> {\centering\arraybackslash}p{1.3cm} |>{\centering\arraybackslash}p{1.8cm} |>{\centering\arraybackslash}p{1.5cm}|>{\centering\arraybackslash}p{1.8cm}|>{\centering\arraybackslash}p{1.7cm}| >{\centering\arraybackslash}p{1.9cm}|}
\hline
\textbf{Table Name} & \textbf{Machine Event} & \textbf{Machine Attribute} & \textbf{Job Event} & \textbf{Task Event} & \textbf{Task Constraint} & \textbf{Task Usage}\\ [0.5ex] % inserts
\hline \hline
Nbr of Files & 1 & 1 & 500 & 500 & 500 & 500  \\\hline
Nbr Data entries & 37,780 & 10,748,566 & 2,012,242 & 144,648,288 & 28,485,619 & 1,232,792,102\\\hline
Avg Entries/file & 37,780 & 10,748,566 & 4,024.5 & 289,296.6 & 56,971.2 & 2,465,584.2\\\hline
Nbr of Attributes & 6 & 5 & 8 & 13 & 6 & 19\\\hline
\end{tabular}
\vspace{-15 pt}
\end{table*}

\subsection{Data Extraction and Processing}
Figure~\ref{Figure:overview} describes our proposed methodology to extract and analyse Gloogle traces files. First, we parse the CSV files containing scheduling events and the resources usage of tasks and jobs. Then we extract the attributes describing the tasks/jobs. %and help to identify/profile the failed ones.
Next, we map the failure events of tasks to the failure events of jobs, to identify correlations between them. The remainder of this section elaborates more on each of these steps.

\subsubsection{Extraction of Tasks/Jobs Attributes}
We implemented a Java program to parse task (respectively job) events and usage files and extract useful attributes. For each task, we extract the following metrics: job ID; task ID; waiting time; service time; scheduling class; priority; requested and used \textit{CPU}, \textit{RAM} and \textit{Disk Space}; number of previous dependent tasks that were finished, killed, failed, evicted, lost or unscheduled; number of times the task was rescheduled after being failed and the final status of the task.
For each job, we extract the: job ID; waiting time; service time; scheduling class; number of finished, killed, failed, evicted, lost or unscheduled tasks within this job; total number of tasks composing the job and the final status of the job. Table~\ref{Table:Jobs:Tasks:Description} presents a description of the attributes and our rationale for selecting them. %as shown in Table~\ref{Table:Jobs:Tasks:Description}.
\begin{table}[!h]
\caption{Jobs and Tasks Attributes}
\label{Table:Jobs:Tasks:Description}
\centering \scriptsize
\begin{tabular} {|p{3cm} |p{6cm}|p{5cm}| }  %{| >{\centering\arraybackslash}p{3cm} | >{\centering\arraybackslash}p{5cm}|}
\hline
\textbf{Jobs/Task Attributes} & \textbf{Description} & \textbf{Rationale}\\  % inserts
\hline \hline
Job ID & Immutable and unique identifier for a job & Used to identify a job\\ \hline
Task ID & Immutable and unique identifier for a task & Used to identify a task \\ \hline
Waiting Time & Amount of time from being submitted until being assigned to a machine & Used to characterize scheduling delay\\ \hline
Service Time  & Amount of time from being assigned to a machine until being finished/failed& Used to capture the execution time\\ \hline
Scheduling Class & Latency-sensitivity type of a Job/Task & It represents a local machine policy for accessing resources\\ \hline
Priority & Preemption type of a task & Used to capture the priority of a task/job when accessing resources \\ \hline
Nbr Finished, Killed, Failed, & Number of finished, killed, failed, evicted, & Used to capture the proportion of tasks\\  %Describe distribution of tasks within a job \\
Evicted, Lost and Unscheduled tasks &   lost and unscheduled tasks within a job &  within a job, that are evicted, lost, or unscheduled \\ \hline
Nbr Pre. Finished, Killed, Failed, Evicted, Lost and Unscheduled tasks & Number of previous finished, killed, failed, evicted, lost and unscheduled dependent tasks for a task& Used to capture failure events that are dependent on a task \\ \hline %before processing it \\ \hline
Nbr of Reschedule Events& Number of reschedule events of failed task & Used to capture the number of times that a failed task was rescheduled \\ \hline %Describes how many times the tasks was failed then rescheduled \\ \hline
Total Nbr of Tasks of Job & Total number of tasks within each job & Used to capture the distribution of tasks within the jobs \\ \hline
Requested/Used CPU & Amount of requested/used CPU for a task & \multirow{3}*{Used to capture the} \\  %\hline
\cline{1-2} Requested/Used RAM & Amount of requested/used RAM for a task &  allocation and usage of ressources\\  %\hline
\cline{1-2}Requested/Used Disk & Amount of requested/used Disk for a task &  \\  \hline
Final Status  &  Final state on a scheduling life-cycle & Used to describe the processing outcome of a task/job\\
\hline
\end{tabular}
\label{table:nonlin}
\vspace{-15pt}
\end{table}
\subsubsection{Identification/Profiling of the Failed Tasks/Jobs}\label{taskstatus}
To identify failed tasks/Jobs we look for one of the following status :
failed, killed, lost, evicted and unscheduled. Tasks/jobs with a dependent task/job that was failed is consider to be failed.
If there are some missing information in the files about a task/job's final status, we consider that the tasks/job is lost.

\subsubsection{Mapping between Failed Tasks and Jobs}
Since jobs are composed of one or multiple tasks.
We extracted the distribution of tasks within each job according to their final status {finished, failed, killed, evicted, lost and unscheduled} to analyse the correlation between tasks scheduling outcomes and job scheduling outcome.

%% file: model.tex
\section{Case Study Results}
\label{sec:casestudy:results}
This section presents and discusses the results of our first two research questions:
\vspace{-0.1cm}
\subsection*{\textbf{RQ1:} \RQone}
\subsubsection*{\textbf{Motivation}} This question is preliminary to the others (\ie{} \textbf{RQ2} and \textbf{RQ3}). It aims to examine the proportion of failed, killed, evicted, lost, and unscheduled jobs that occurred in Google clusters over a period of one month. If these events are very frequent, then they are worth studying in more details. We also examine the waiting and service times of jobs (respectively tasks) in each category to evaluate the impact of task and job failures on processing costs.% in the cluster.
% , in order to predict task and jobs failures and eventually prevent their occurrence.
\subsubsection*{\textbf{Approach}} We address this question by extracting information about unfinished tasks and jobs from our data set following the method described in Section~\ref{taskstatus}. We used all jobs files. However, we experienced very long processing times (\ie{} lasting multiple days) when analysing the task files. We decided to reduce the amount of data to process by randomly sampling 2\% (10 files out of 500) of the tasks files, in order to speed up our analysis. However, we verified the relevance of our sample by re-sampling the data-set multiple times (i.e. 5 times) and comparing the results of our analysis.

\subsubsection*{\textbf{Findings}}
\begin{figure}[ht]
        \centering
    \begin{subfigure}[b]{0.35\linewidth}        %% or \columnwidth
        \centering
        \includegraphics[width=\linewidth]{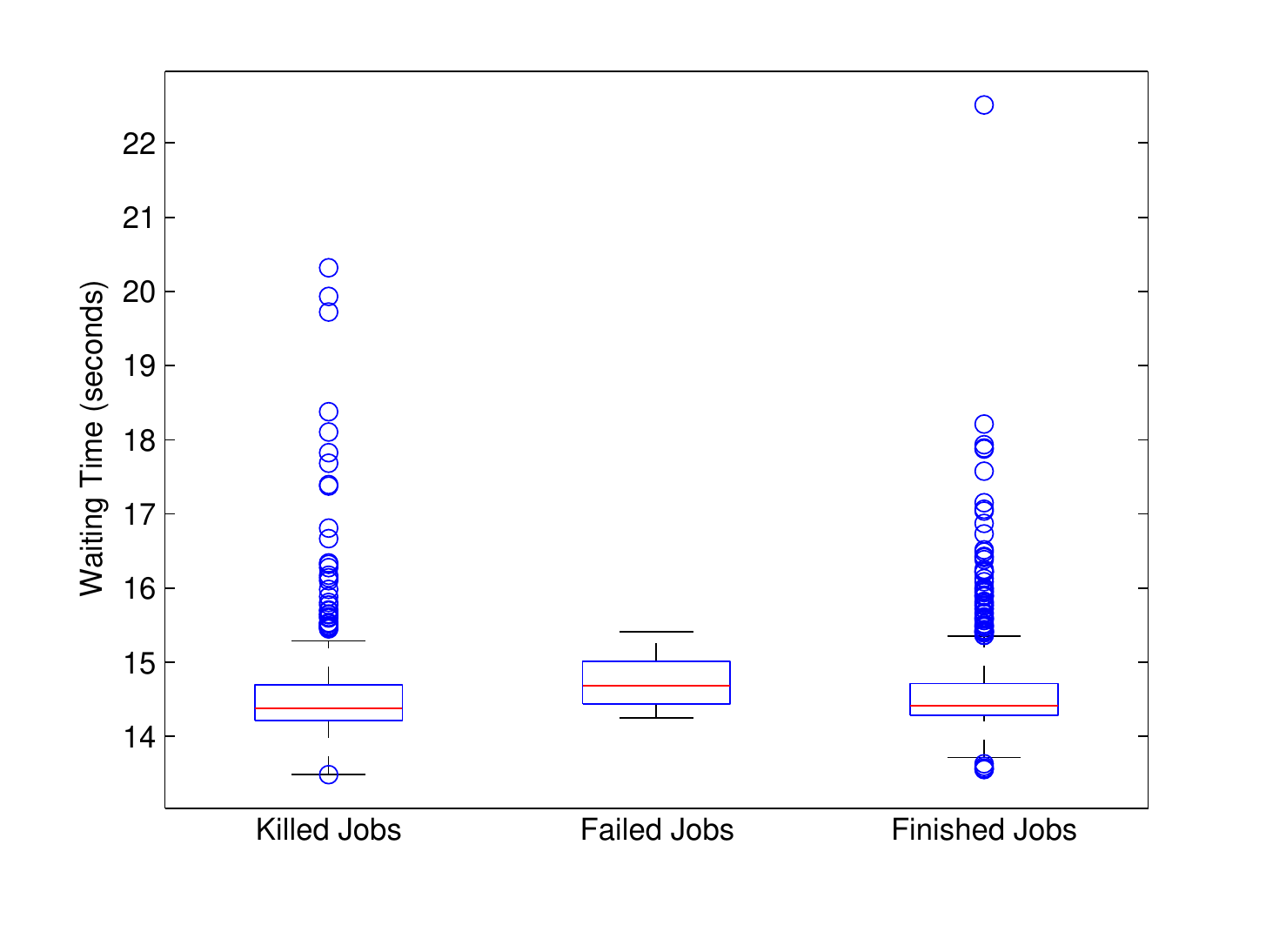}
        \vspace{-5pt}
        \caption{Waiting Time of Jobs}
        \label{Figure:WaitingTimeJobs}
    \end{subfigure}
    \begin{subfigure}[b]{0.35\linewidth}        %% or \columnwidth
        \centering
        \includegraphics[width=\linewidth]{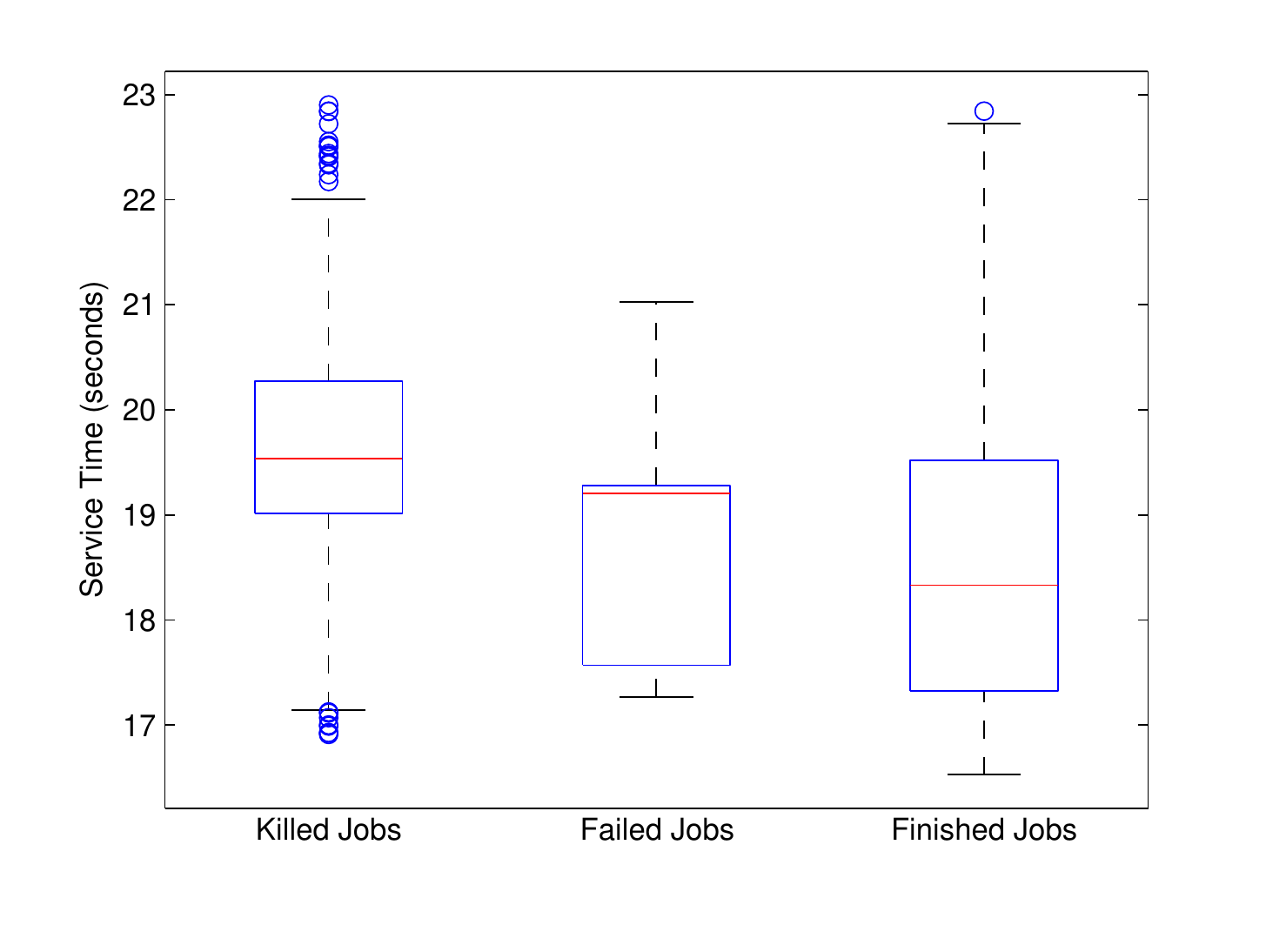}
        \vspace{-5pt}
        \caption{Service Time of Jobs}
        \label{Figure:ServiceTimeJobs}
    \end{subfigure}
        \caption{Waiting and Service Time of Jobs (log scale)}
        \label{Figure:WaitingServiceTimeTasks}
\vspace{-5pt}
\end{figure}

%\textbf{Job Level} : we noticed that
\textbf{Only \textbf{58.47\%} of the submitted jobs were finished successfully and the rest were killed, failed, unscheduled or evicted as shown in Table \ref{Table:Distribution:Jobs:Traces}.} We also observed that few jobs (\ie{} 0.8\%) were not scheduled. Also, the number of killed task is very important (almost 40\%) compared to the finished ones.
\begin{table*}[ht]
\caption{Distribution of Jobs across Google Traces Files}
\label{Table:Distribution:Jobs:Traces}
\centering \scriptsize
\begin{tabular}{| >{\centering\arraybackslash}p{3cm} | >{\centering\arraybackslash}p{2cm}| >{\centering\arraybackslash}p{2cm}|}
\hline
Job Status & Nbr Jobs & \%\\ [0.5ex]
\hline\hline
Finished & 379586 & 58.47\% \\\hline
Killed   & 255280 & 39.33\% \\\hline
Failed   & 9080   & 1.4\%   \\\hline
Evicted  & 14     & 0.0\%   \\\hline
Lost     & 0      & 0.0\%   \\\hline
Unscheduled & 5169 & 0.8\% \\
\hline\hline
Total & 649129    & 100\% \\ \hline
\end{tabular}
\end{table*}
Furthermore, we also found that failed and killed jobs are characterized by long waiting times, as described in Figure~\ref{Figure:WaitingTimeJobs}. Meaning that a reduction of the amount of failed and killed jobs can help reduce processing times on clusters, which would result in energy and resources savings. % (a reduction of scheduling delays may increase the success rate of scheduled jobs. %, which can be explained by difficulties in the scheduler to reduce the scheduling delay and allocate the required resources.
We also noticed that killed and failed jobs have longer service time compared to other finished jobs, as shown in Figure~\ref{Figure:ServiceTimeJobs}. %which may be explained by difficulties to satisfy the requirements of these jobs.
Therefore, it is very important to identify the main reasons that lead to jobs failure in order to reduce their processing cost (in terms of service and waiting times) and consequently improve the cluster performance.

% First, we noticed that the obtained statistics from different samples are so close and give almost the same distribution of tasks across the sampled files.}
%Consequently, we decided to use only on sample 10 files fromto analyse the scheduling outcomes of tasks while minimizing their processing time.\\

\begin{figure}[ht]
    \centering
    \begin{subfigure}[b]{0.35\linewidth}        %% or \columnwidth
        \centering
        \includegraphics[width=\linewidth]{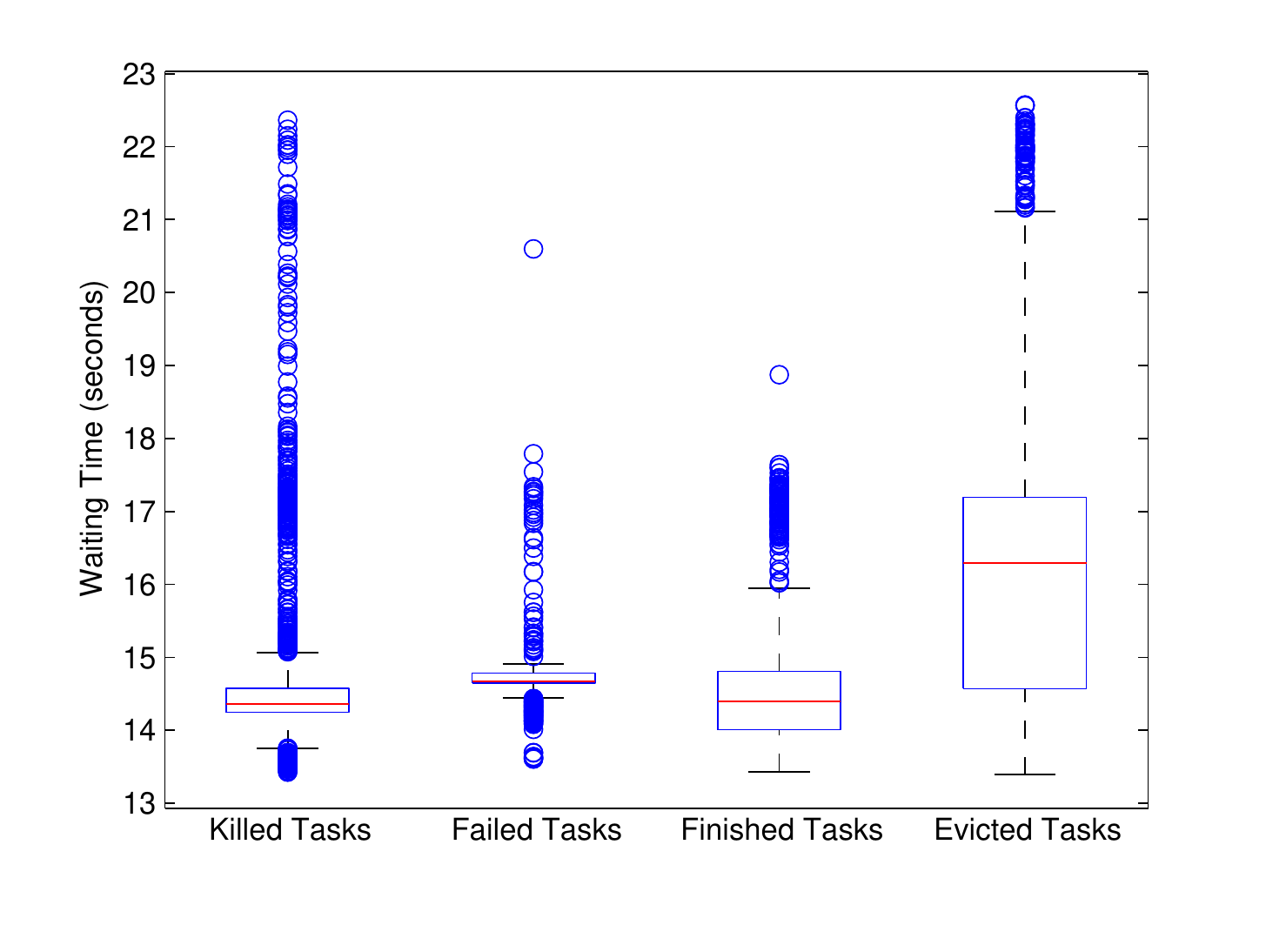}
        \vspace{-5pt}
        \caption{Waiting Time of Tasks}
        \label{Figure:WaitingTimeTasks}
    \end{subfigure}
    \begin{subfigure}[b]{0.35\linewidth}        %% or \columnwidth
        \centering
        \includegraphics[width=\linewidth]{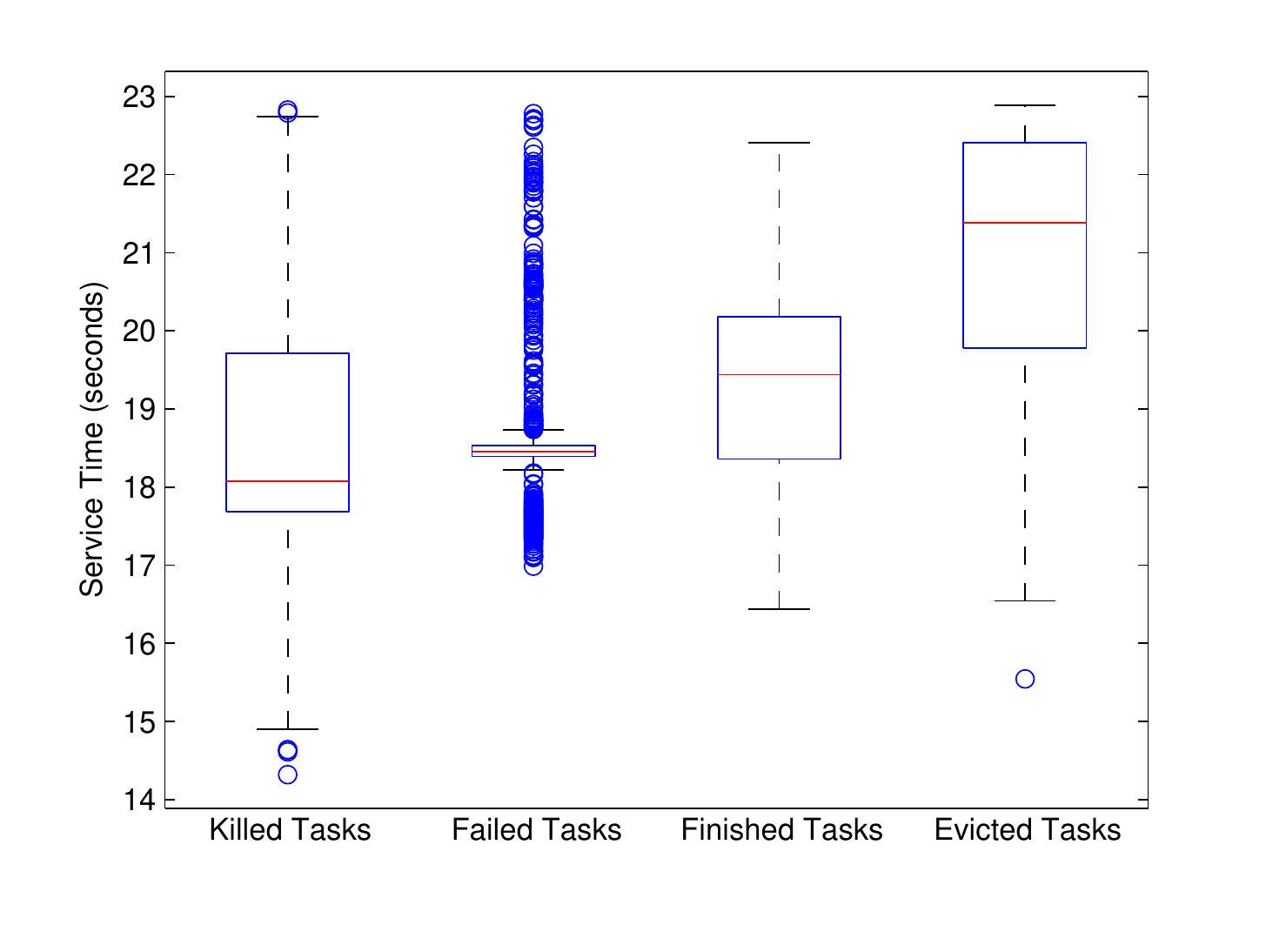}
        \vspace{-5pt}
        \caption{Service Time of Tasks}
        \label{Figure:ServiceTimeTasks}
    \end{subfigure}
    \caption{Waiting and Service Time of Tasks (log scale)}
        \label{Figure:WaitingServiceTimeTasks}
\vspace{-5pt}
\end{figure}

\begin{table}[ht]
\caption{Distribution of Task across 10 Google Traces Files}
\label{Table:Distribution:Task:Traces}
\centering \scriptsize
\begin{tabular}{| >{\centering\arraybackslash}p{3cm} | >{\centering\arraybackslash}p{2cm}| >{\centering\arraybackslash}p{2cm}|}
\hline
Task Status & Nbr Tasks & \%\\ [0.5ex]
\hline\hline
Finished & 33020 & 52.22\% \\\hline
Killed   & 8473 & 13.40\% \\\hline
Failed   & 7044   & 11.14\%   \\\hline
Evicted  & 12798     & 20.24\%   \\\hline
Lost     & 0      & 0.0\%   \\\hline
Unscheduled & 1897 & 3\% \\
\hline\hline
Total & 63234    & 100\% \\ \hline
\end{tabular}
\vspace{-8pt}
\end{table}
\textbf{More than \textbf{24\%} of tasks were failed or killed and only \textbf{52\%} of tasks were finished successfully. However, 97\% of tasks were scheduled successfully, \ie{} only 3\% of tasks were not scheduled and were resubmitted for scheduling (some tasks were resubmitted up to 182 times)}. We also observed a high percentage of evicted tasks (\ie{} 20\%) as shown in Table~\ref{Table:Distribution:Task:Traces}. Evicted tasks had lower priority compared to production and monitoring tasks. In addition, we also observed that evicted tasks have long waiting time and service time compared to other tasks (see Figure~\ref{Figure:WaitingTimeTasks} and Figure~\ref{Figure:ServiceTimeTasks}). Also, failed and killed tasks are characterized by long waiting and execution time. Therefore, it is crucial to reduce the amount of failed and evicted tasks if we want to optimize jobs and tasks processing times (jobs are composed of multiple tasks). We repeated the analysis on all the other samples collected from the Google trace data and obtain similar results; suggesting that the observed high rates of job and task failures are not specific to the studied sample but rather likely representative of the general situation of jobs and tasks scheduling issues in Google clusters. %We obtained the same above mentioned results on all the other samples collected from the Google trace data set which suggests that our sample . \textcolor{red}{Hence we We noticed that we can extrapolate these obtained results on the rest of task files and the 2\% of tasks files are representative for the analysis and can reduce the total computation time.}
%n order to cost related to the processing of these tasks by predicting them and reducing the task failure events.

\subsection*{\textbf{RQ2:} \RQtwo}
\subsubsection*{\textbf{Motivation}}
In \textbf{RQ1}, we observed that scheduled tasks and jobs experience high failure rates. In this research question, we examine the correlation between the characteristics of tasks (respectively jobs) and scheduling outcomes, in order to predict tasks failures and eventually prevent their occurrence. %We also investigate strategies to predict the scheduling outcome of task and jobs.
Specifically, our goal is to determine whether the scheduling outcome of a task can be predicted early on before it actually happens. Such predictions can be used to reschedule potential failing tasks quickly on appropriate clusters with adequate resources in order to ensure their timely and successful completion.

\subsubsection*{\textbf{Approach}}
We extract all the metrics described in Table~\ref{Table:Jobs:Tasks:Description}. We use the Spearman rank correlation~\cite{cortina2012handbook} to test the association between these metrics and task scheduling outcomes. %and filter out all correlated metrics, using .
We also perform a Variance Inflation Factor (VIF) analysis to examine multi-collinearity between the metrics. We use a threshold of 5 to decide on the multi-collinearity of the metrics, \ie{} metrics with VIF result greater than 5 are considered as correlated. We choose several regression and classification algorithms in \textit{R}~\cite{[23]} to build models: GLM (General Linear Model), Random Forest, Neural Network, Boost, Tree and CTree (Conditional Tree). GLM is an extension of linear multiple regression for a single dependent variable. It is extensively used in regression analysis. Boost creates a succession of models iteratively, each model being trained on a data set in which points misclassified by the previous model are given more weight. All the successive models are weighted according to their success and their outputs are combined using voting or averaging to create a final model. Neural networks are graphs of interconnected nodes organised in layers. The predictors (or inputs) form the bottom layer, and the forecasts (or outputs) form the top layer. Decision Tree is a widely used classification approach to predict a binary result. CTree is a different implementation of Decision Tree. Based on Decision Tree, Leo Breiman and Adele Culter developed Random Forest, which uses a majority voting of decision trees to generate classification (predicting, often binary, class label) or regression (predicting numerical values) results~\cite{breiman2001random}. Random Forest offers good out-of-the-box performance and has performed very well in different prediction benchmarks~\cite{mende2010effort}. The algorithm yields an ensemble that can achieve both low bias and low variance~\cite{diaz2006gene}.
We use different training and testing data sets for both jobs and tasks. We apply 10-fold random cross validation to measure the accuracy, the precision, and recall of the prediction models \cite{[15]}. The accuracy is $\frac{TP + TN}{TP+TN+FP+FN}$, the precision is $\frac{TP}{TP+FP}$, and the recall is $\frac{TP}{TP+FN}$, where $TP$ is the number of true positives, $TN$ is the number of true negatives, $FP$ is the number of false positives, and $FN$ is the number of false negatives. %can be defined as the proportion of true prediction (valid + wrong) among the total number of examined tests. Precision is the the proportion of the valid true results against all the true results to measure the relevance of the prediction. The recall can determine the number of prediction among the obtained ones.}
In the cross validation, each data set is randomly split into ten folds. Nine folds are used as the training set, and the remaining one fold is used as the testing set.
\subsubsection*{\textbf{Findings}}
\paragraph{Job Level}
We analysed the correlation between job attributes (captured by the metrics from Table~\ref{Table:Jobs:Tasks:Description}) and jobs scheduling outcomes. We observed multi-collinearity between the following attributes: total number of tasks, service and waiting time and number of unscheduled/lost task (\ie{} VIF results were over 5). We found a strong correlation between the number of finished, failed, killed and evicted tasks within a job (with VIF values of respectively 1.87, 4.17, 3.65 and 2.42) and the final status of the job. Therefore we conclude that, when there are dependencies between the tasks composing a job, the scheduling outcome of the job is impacted by the scheduling outcome of its contained tasks. For example, Rule 1 which was among the rules obtained using the Random Forest algorithm shows the relation between the scheduling outcome of a job and the scheduling outcome of its contained tasks. %) We also used the decision Tree model to extract the a rules to characterize jobs scheduling outcomes described as follow:
\begin{table}[ht]
\label{Table:Algo1}
\centering \scriptsize
\begin{tabular}{|l| }
   \hline
Rule 1: Relation Between the Scheduling Outcomes of Jobs and Tasks \\ \hline \hline
\textbf{if} (number killed tasks \textless  0.5) \textbf{then} \\
\hspace*{0.3cm} \textbf{if} (number finished tasks \textless  0.5) \textbf{then} \\
\hspace*{0.5cm}     Failed \\
\hspace*{0.3cm}   \textbf{else} \\
\hspace*{0.5cm}      Finished  \\
\hspace*{0.3cm}   \textbf{end if} \\
\textbf{else}  \\
\hspace*{0.3cm}   Failed \\
\textbf{end if }     \\ \hline
\end{tabular}
\vspace{-7pt}
\end{table}

\begin{figure}[ht]
    \centering
    \begin{subfigure}[b]{0.35\linewidth}        %% or \columnwidth
        \centering
        \includegraphics[width=60mm,height=50mm]{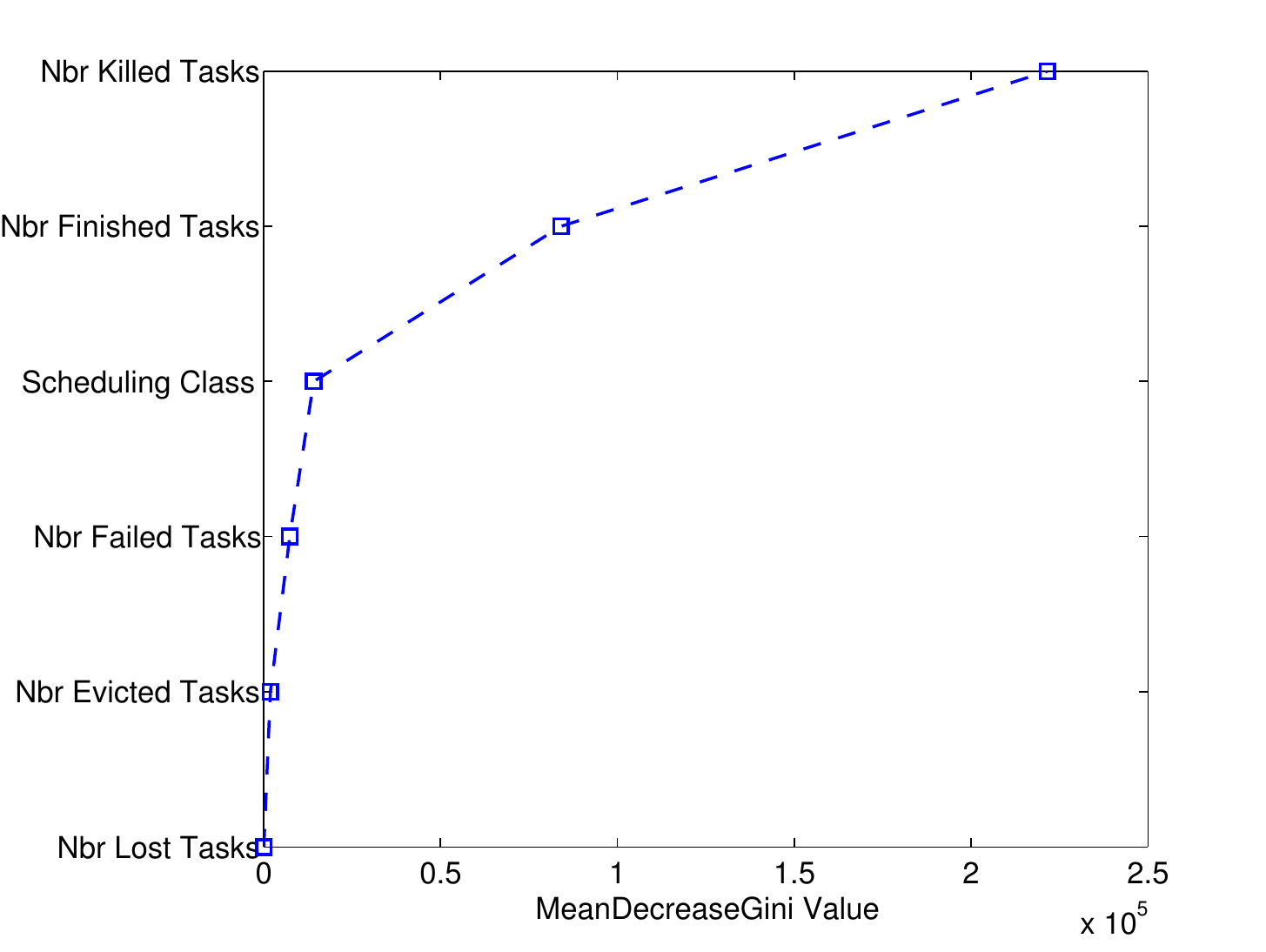}
        \vspace{-15pt}
        \caption{Jobs Attributes}
        \label{Figure:MeanDecreaseGiniJobs}
    \end{subfigure}
    ~
    \begin{subfigure}[b]{0.35\linewidth}        %% or \columnwidth
        \centering
        \includegraphics[width=70mm,height=50mm]{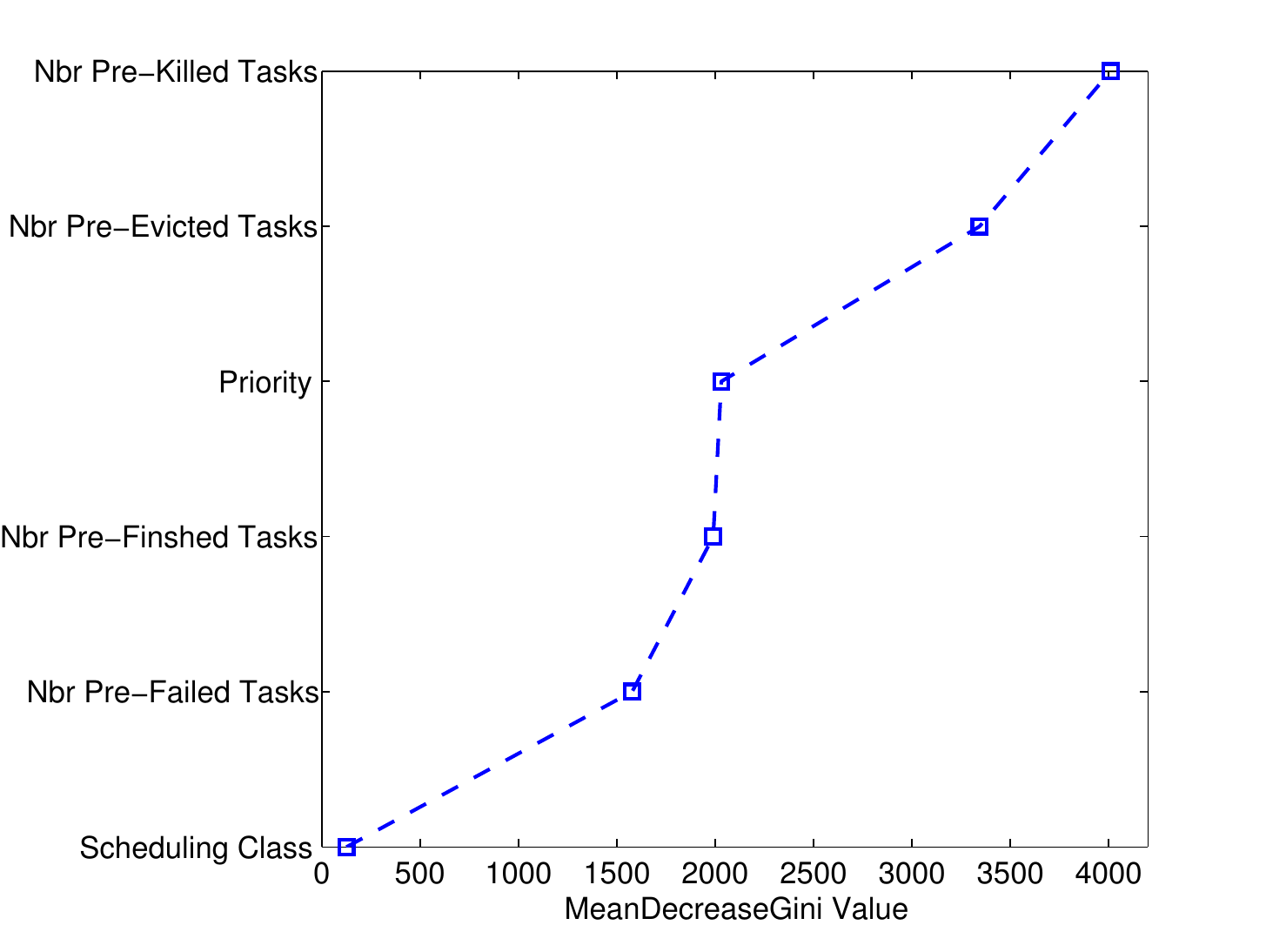}
        \vspace{-15pt}
        \caption{Task Attributes}
        \label{Figure:MeanDecreaseGiniTasks}
    \end{subfigure}
    \caption{Importance of Jobs/Tasks Attributes using Random Forest}
        \label{Figure:MeanDecreaseGini}
\vspace{-5pt}
\end{figure}

\begin{table}[ht]
\caption{Accuracy, Precision, Recall (In \%) obtained from different Algorithms: (Random 10-fold Cross Validation)}
%Tree, Random Forest, Neural Network, Boost, Glm and CTree: (Random K=10)} % title of Table
\centering \scriptsize
\label{Table:ResultsJobs:R}
\begin{tabular}{|>{\centering\arraybackslash}p{1.5cm}|> {\centering\arraybackslash}p{3cm} |>{\centering\arraybackslash}p{1.5cm} |>{\centering\arraybackslash}p{1.5cm}|>{\centering\arraybackslash}p{1.5cm}|}
\hline
\textbf{\# File} & \textbf{Algo. }&  \textbf{Acc.}  & \textbf{Pre.} & \textbf{Rec.} \\ [0.5ex] % inserts table
\hline\hline
\multirow{6}{*}{{\bf 10}} & Tree           & 66.7       & 83.7       & 66.7       \\ \cline{2-5} 
                          & Boost          & 75         & 89.1       & 75         \\ \cline{2-5} 
                          & Glm            & 68.8       & 90.7       & 61.7       \\ \cline{2-5} 
                          & CTree          & 61.8       & 89.7         & 54.9       \\ \cline{2-5} 
                          & Random Forest  & {\bf 85.6} & {\bf 94.2} & {\bf 85.9} \\ \cline{2-5} 
                          & Neural Network & 56         & 67.6       & 66.6       \\ \hline
\end{tabular}
\label{table:nonlin}
%\vspace{-15pt}
\end{table}
\textbf{Random Forest achieves the best precision and recall when predicting the outcome of job scheduling. It can achieve an average accuracy of \textbf{85.6\%}, a precision of 94.2\% and a recall of 85.9\%.} Table~\ref{Table:ResultsJobs:R} summarises the performance of the six models.
Only 10 job files mapped to the tasks files were used in the construction of the models because of our processing resources limitation. In fact, the 10 files contained 2, 594 jobs and 6, 3234 tasks which processing took 100 minutes (processing 10 cross-validations on one file took only 9.5 minutes).\\
Figure~\ref{Figure:MeanDecreaseGiniJobs} shows the importance of each attributes used in our Random Forest model. According to the Mean Decrease Gini score, the most important attributes are ordered as follow: \textit{Nbr Killed tasks, Nbr Finished tasks, Scheduling.Class, Nbr Failed tasks, Nbr Evicted tasks, Nbr Lost tasks}. This result is consistent with Rule 1 (in which number of killed and finished tasks are shown to impact the scheduling outcome of jobs).\\

%In this context, we noticed that the obtained association rules reflects almost the same obtained results as shown in~Figure\ref{Figure:MeanDecreaseGiniJobs} where the main jobs attributes that can affect the scheduling outcome are: number of killed and finished tasks within a job.
\paragraph{Tasks Level}
We analysed the relation between task attributes and scheduling outcomes and obtained a strong correlation between the number of previously finished, killed, failed and evicted tasks, priority and the scheduling outcome of tasks (having respectively these VIF values: 1.14, 1.02, 1.07, 1.03, 1.06). We observed multi-collinearity between the number of rescheduled tasks, service time, waiting time, and the amount of requested/used resources (\textit{CPU, RAM, Disk}). In addition, we noticed that the resources assigned to each task were higher than the requested resources (which can be explained by the overbooking strategy followed by Google~\cite{[16]}). Overall, tasks characterized by dependent tasks that failed in the past have a high probability to fail in the future, as shown on Rule 2 (obtained using the Random Forest algorithm). Also, tasks with low priority values have a high probability to be evicted~\cite{[11]}.\\
\begin{table}[ht]
\label{Table:Algo2}
\centering \scriptsize
\begin{tabular}{|l| }
 \hline
Rule 2 : Relation between Tasks Attributes Scheduling Outcomes  \\ \hline \hline
\textbf{if} (number pre-killed tasks  \textgreater   0.5) \textbf{then} \\
\hspace*{0.3cm}  Failed \\
\textbf{else} \\
\hspace*{0.3cm} \textbf{if} (priority \textless 1.5) \textbf{then} \\
\hspace*{0.7cm}   \textbf{if}(number of pre-evicted tasks  \textgreater  0.5) \textbf{then}\\
\hspace*{0.9cm}      Failed \\
\hspace*{0.7cm}    \textbf{else} \\
\hspace*{0.9cm}      \textbf{if}(number of pre-failed tasks \textgreater  0.5) \textbf{then} \\
\hspace*{1cm}       Failed \\
\hspace*{0.9cm}      \textbf{else} \\
\hspace*{1cm}        Finished \\
\hspace*{0.9cm}     \textbf{end if} \\
\hspace*{0.7cm}   \textbf{end if }\\
\hspace*{0.3cm} \textbf{else} \\
\hspace*{0.7cm}  Finished \\
\hspace*{0.3cm} \textbf{end if }   \\
\textbf{end if }     \\ \hline
\end{tabular}
\vspace{-7pt}
\end{table}

When predicting failed tasks, Random Forest can achieve an average accuracy of \textbf{95.8\%}, a precision of \textbf{97.4\%} and a recall of \textbf{96.2\%} (see Table~\ref{Table:ResultsTasks:R}).
From Table~\ref{Table:ResultsTasks:R}, we also observe that results obtained with 1 file are quite similar to the results obtained with 10 files. We explain this result by the fact that the distribution of failure events in these files are very similar (as shown in \textbf{RQ1}). %are so close because of the same failure events distribution in the tasks files. So, we got approximately the same results from different files. Thus, we adopted the same approach as well as for the job failure prediction and we used some of the files while predicting the failure events to reduce the execution time of this algorithms. Also, we noticed that the Random Forest performs better that other predictions algorithms.
\begin{table}[ht]
\caption{ Accuracy, Precision, Recall (In \%) obtained from from different Algorithms: (Random 10-fold Cross Validation)}
%Tree, Random Forest, Neural Network, Boost, Glm and CTree: Random K-Cross Validation (k=10)}
\label{Table:ResultsTasks:R}
\centering \scriptsize
\begin{tabular}{|>{\centering\arraybackslash}p{3cm}|> {\centering\arraybackslash}p{0.9cm}|> {\centering\arraybackslash}p{0.9cm}|> {\centering\arraybackslash}p{0.9cm}|> {\centering\arraybackslash}p{0.9cm}|> {\centering\arraybackslash}p{0.9cm}|> {\centering\arraybackslash}p{0.9cm}|}
\hline
\multirow{2}{*}{{\bf Algo.}} & \multicolumn{2}{c|}{{\bf Acc.}} & \multicolumn{2}{c|}{{\bf Pre.}} & \multicolumn{2}{c|}{{\bf Rec.}} \\ \cline{2-7} 
                             & {\bf 1 F.\textsuperscript{*}}   & {\bf 10 F.}  & {\bf 1 F.}   & {\bf 10 F.}  & {\bf 1 F.}  & {\bf 10 F.}  \\ \hline \hline
Tree                         & 74             & 66.2           & 84.8           & 77             & 74.3          & 66.7            \\ \hline
Boost                        & 88.6           & 89.3           & 99.5           & 99.6           & 80.8          & 81.4            \\ \hline
Glm                          & 70.8           & 74.5           & 97             & 99.6           & 52.6          & 55.2            \\ \hline
CTree                        & 87.4           & 92.5           & 94.9           & 98             & 85.6          & 98.2            \\ \hline
Random Forest                & {\bf 95.8}     & {\bf 97.3}     & {\bf 97.4}     & {\bf 98.1}     & {\bf 96.2}    & {\bf 97.7}      \\ \hline
Neural Network               & 50             & 50             & 56.4           & 50             & 50            & 50              \\ \hline
\end{tabular}
\begin{tablenotes}
\footnotesize
\item   \textsuperscript{*} F. = File
  \end{tablenotes}
\end{table}
Furthermore, we applied the \textit{MeanDecreaseGini} criteria on the task attributes. We observed that the final status of a task is mainly dependent on some attributes from its historical data including, in order: \textit{Number of Previous Dependent Killed, Failed, Evicted, Finished, Priority, Scheduling Class}, as shown on Figure~\ref{Figure:MeanDecreaseGiniTasks}. This result is consistent with Rule 2 (in which number of previous killed, evicted and failed tasks and the priority of the task, are shown to impact the scheduling outcome of the task).

%% file: results.tex
\section{Results Scheduling Outcomes Prediction: Google Cluster }
\label{sec:predictionresults}
\subsection*{RQ3: \RQthree}
\subsubsection*{\textbf{Motivation}}
Results from \textbf{RQ2} show that a Random Forest model can predict task failure events with high precision (\ie{} 97.4\%) and recall (\ie{}  96.2\%). Therefore, rather than waiting for a scheduled task to fail, a scheduler equipped with such predictions can reschedule the tasks quickly on appropriate clusters with adequate resources. For example by restarting on a different node a task predicted to fail on its current node because of insufficient resources. To quantify the benefits that can be achieved by predicting the scheduling outcome of tasks early, we measure the execution time and numbers of finished tasks and jobs of a scheduler equipped with a Random Forest prediction model. %use the tool kit GloudSim to examine whether the use of prediction models can enable better scheduling decisions.
%
%We aim to prove that our rescheduling strategy using the predictive model can improve the scheduler performance in terms of number of failure events. So, our goal is to evaluate the impact and the benefit of our proposed predictive algorithm on the current scheduling techniques in distributed clusters.
\subsubsection*{\textbf{Approach}}
We used the simulation toolkit \textit{GloudSim} to reproduce the execution traces from the dataset. Indeed, the GloudSim toolkit was developed to simulate the original workload of Google applications in order to support academic research~\cite{[20]}.
We deployed GloudSim using 8 virtual machines (VMs) managed by a \textit{XEN} hypervisor. Each VM had a one \textit{Core(TM)2 Quad CPU} (\ie{} 2.66GHz) and 1024 MB of memory. We implemented a script to collect the following data about submitted tasks : priority, scheduling class, number of previous failure events and requested resources.
%\Foutse{can we list all the data collected instead of using etc}.
We trained the random Forest model from \textbf{RQ2} using historical data generated by GloudSim
%\Foutse{which historical data exactly?}
about scheduled tasks and used it to predict the scheduling outcome of each new task submitted for scheduling.
We used a real-time learning algorithm to update the scheduling policies at fixed time intervals (\ie{} every 10 minutes). Also, our proposed prediction algorithm can be used off-line to add the new learned scheduling rules periodically.
We extend the scheduler implemented in GloudSim to integrate the Random Forest prediction model of tasks. If a scheduled task was predicted to fail, the new scheduler would resubmit the task directly in the scheduling queue without executing it. Only tasks that were predicted to succeed would be processed on the scheduler. On the new scheduler, tasks that are predicted to fail are enqueued until they get a prediction of success. Consequently, if many tasks in a submitted job are predicted to fail, the execution of the job can take a long time and the job may even fail since the tasks will be rescheduled until they are predicted to succeed (which may not occur). %is characterized by a high number of finished tasks it has a higher probability to be completed successfully. On the contrary, if many tasks in a job fail, the job will fail.
We compare the scheduling performance of the new scheduler and the original scheduler implemented in GloudSim, by executing between 100 and 800 tasks and between 100 and 400 jobs. We considered three types of tasks and jobs during the comparison: single (100 tasks-100 jobs), batch (800 tasks-110 jobs) and mix (600 tasks-400 jobs).
The performance of the schedulers were measured in terms of execution times and numbers of finished tasks and jobs. We choose these measures because the execution time of jobs and tasks are two important metrics that capture resource utilisation in the cluster. %the closer it is to the optimal time the better the performance can be improved. In the other hand,
The number of failure events is a good measure of the quality of a scheduler.
%\Foutse{can we explain why we picked these metrics? what they represents?}

\subsubsection*{\textbf{Findings}}
\begin{figure}[ht]
        \centering
        \begin{subfigure}[b]{0.35\linewidth}        %% or \columnwidth
        \centering
        \includegraphics[width=\linewidth]{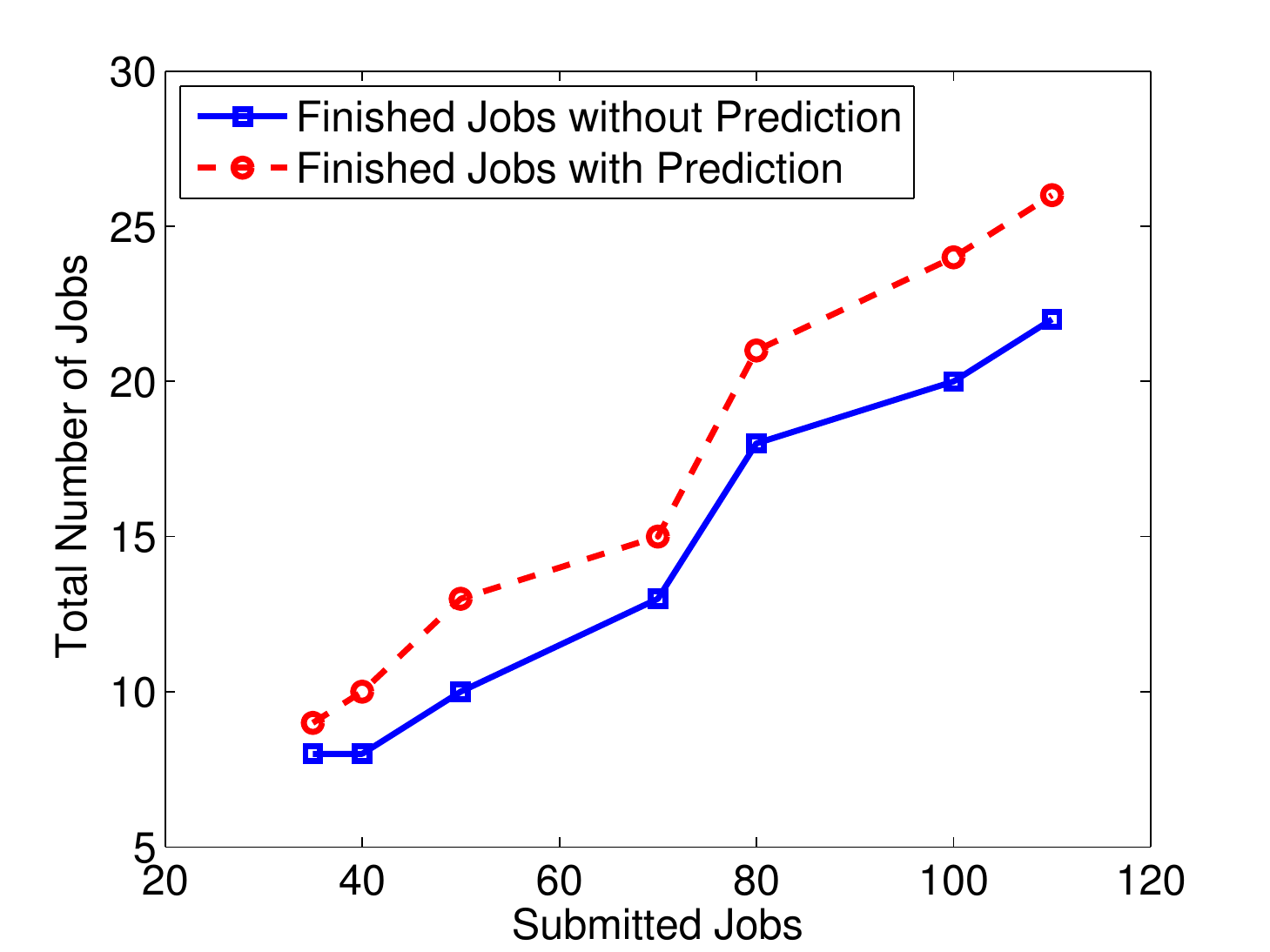}
                \vspace{-5pt}
                \caption{Finished Batch Jobs}
                \label{fig:batchjobsfinished}
        \end{subfigure}
        \begin{subfigure}[b]{0.35\linewidth}        %% or \columnwidth
        \centering
        \includegraphics[width=\linewidth]{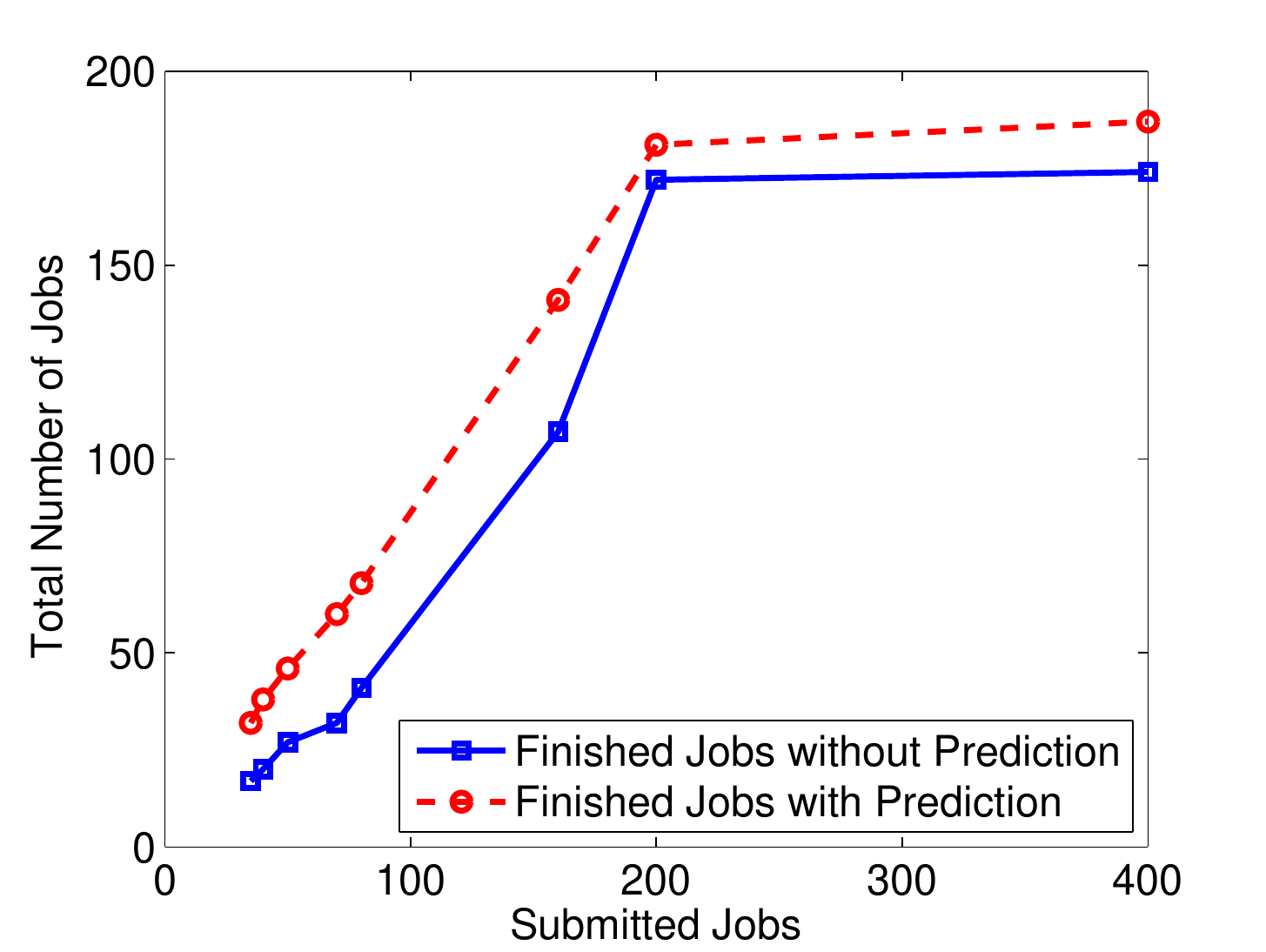}
                \vspace{-5pt}
                \caption{Finished Mix Jobs}
                \label{fig:mixjobsfinished}
        \end{subfigure}%

        \begin{subfigure}[b]{0.35\linewidth}        %% or \columnwidth
        \centering
        \includegraphics[width=\linewidth]{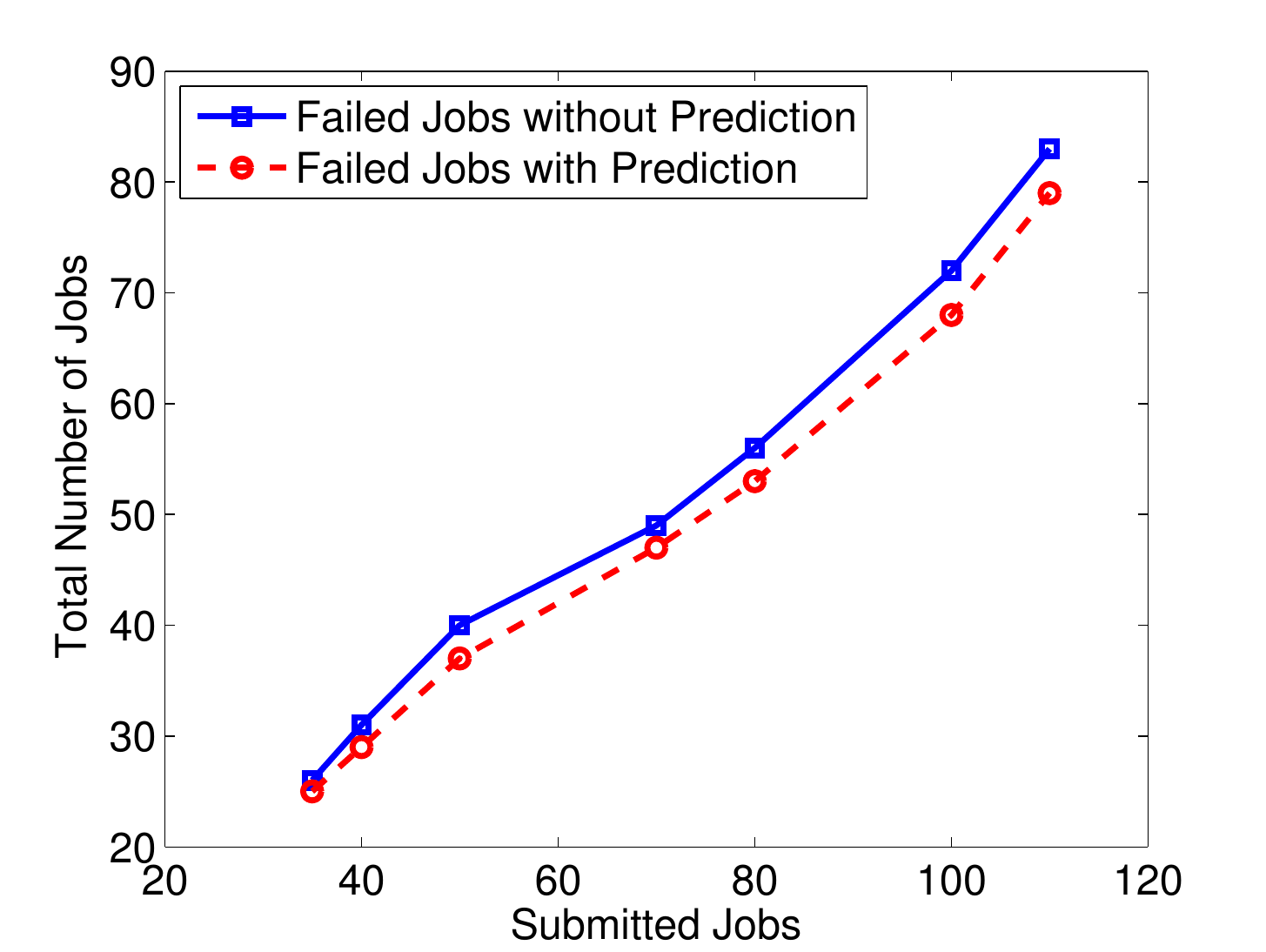}
                \vspace{-5pt}
                \caption{Failed Batch Jobs}
                \label{fig:batchjobsfailed}
        \end{subfigure}
        \begin{subfigure}[b]{0.35\linewidth}        %% or \columnwidth
        \centering
        \includegraphics[width=\linewidth]{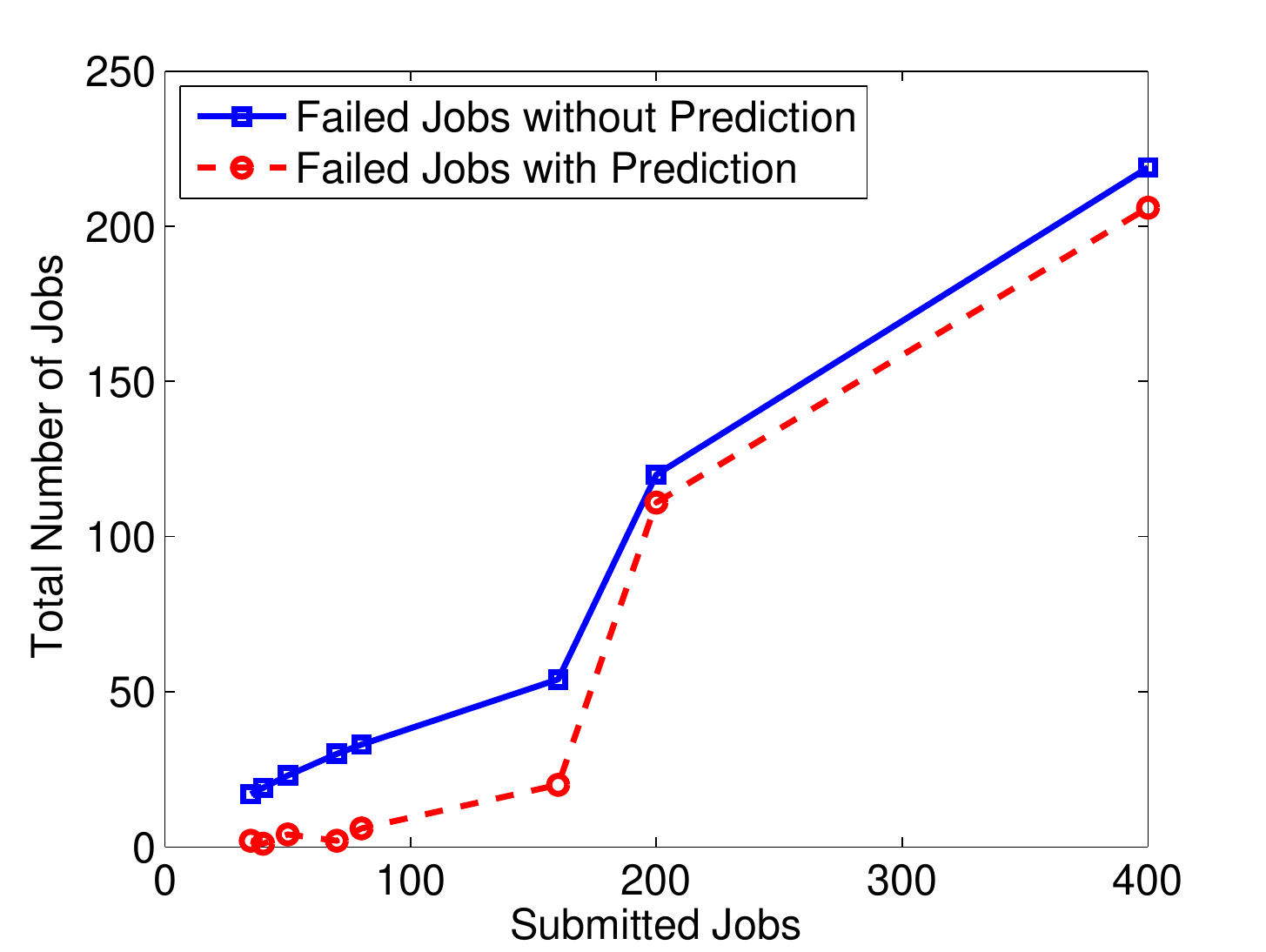}
                \vspace{-5pt}
                \caption{Failed Mix Jobs}
        \label{fig:mixjobsfailed}
        \end{subfigure}%
        \vspace{-5pt}\caption{Distribution of Finished and Failed Jobs}
         \label{fig:failedjobs}
\end{figure}

\begin{figure*}[ht]
        \centering
        \begin{subfigure}[b]{0.3\textwidth}
                \includegraphics[width=5cm,height=40mm]{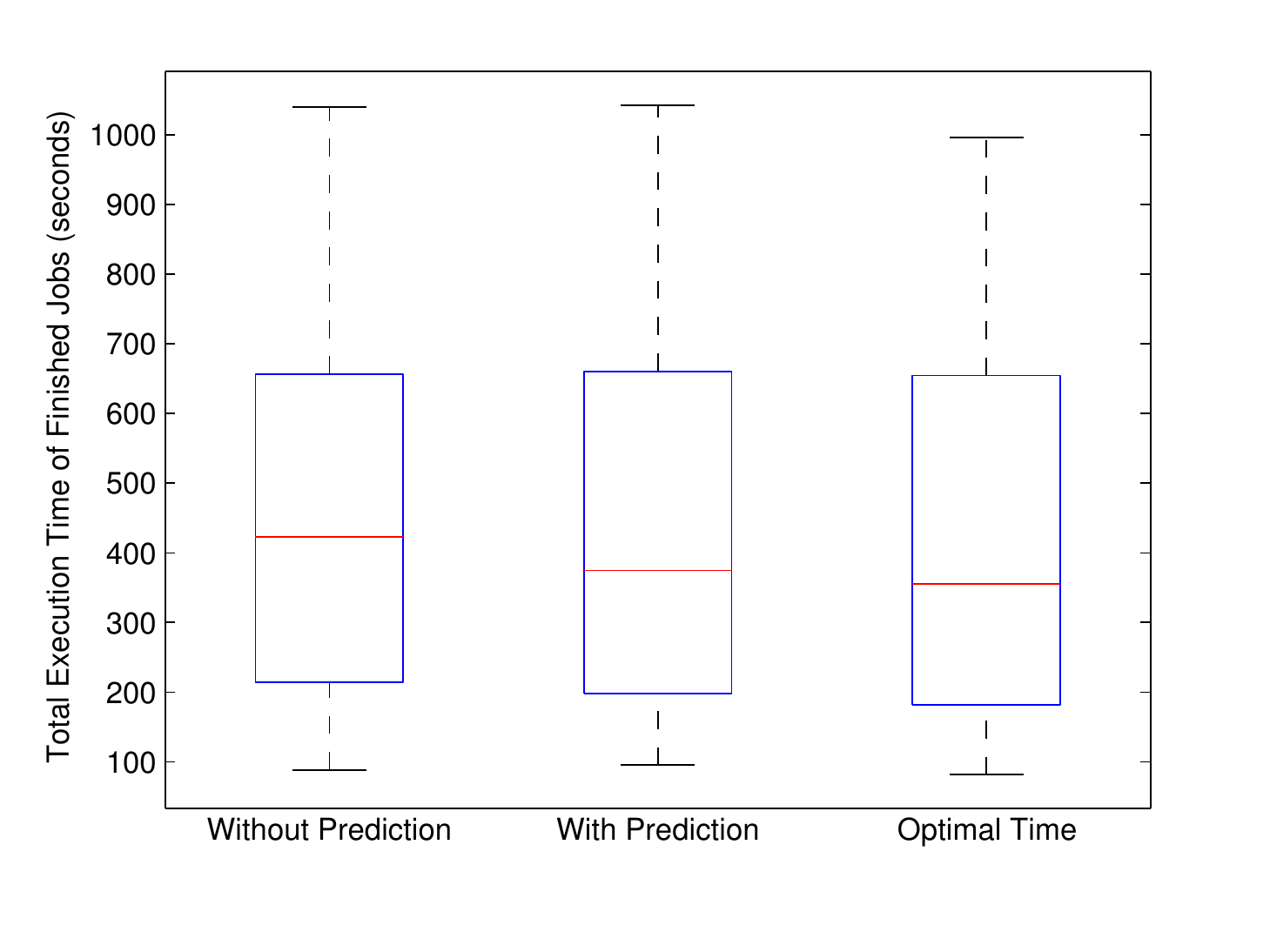}
                \vspace{-7pt}\caption{Single Jobs}
                \label{fig:singlejobs}
        \end{subfigure}
        \vspace{0.00mm}
       \begin{subfigure}[b]{0.3\textwidth}
                \includegraphics[width=5cm,height=40mm]{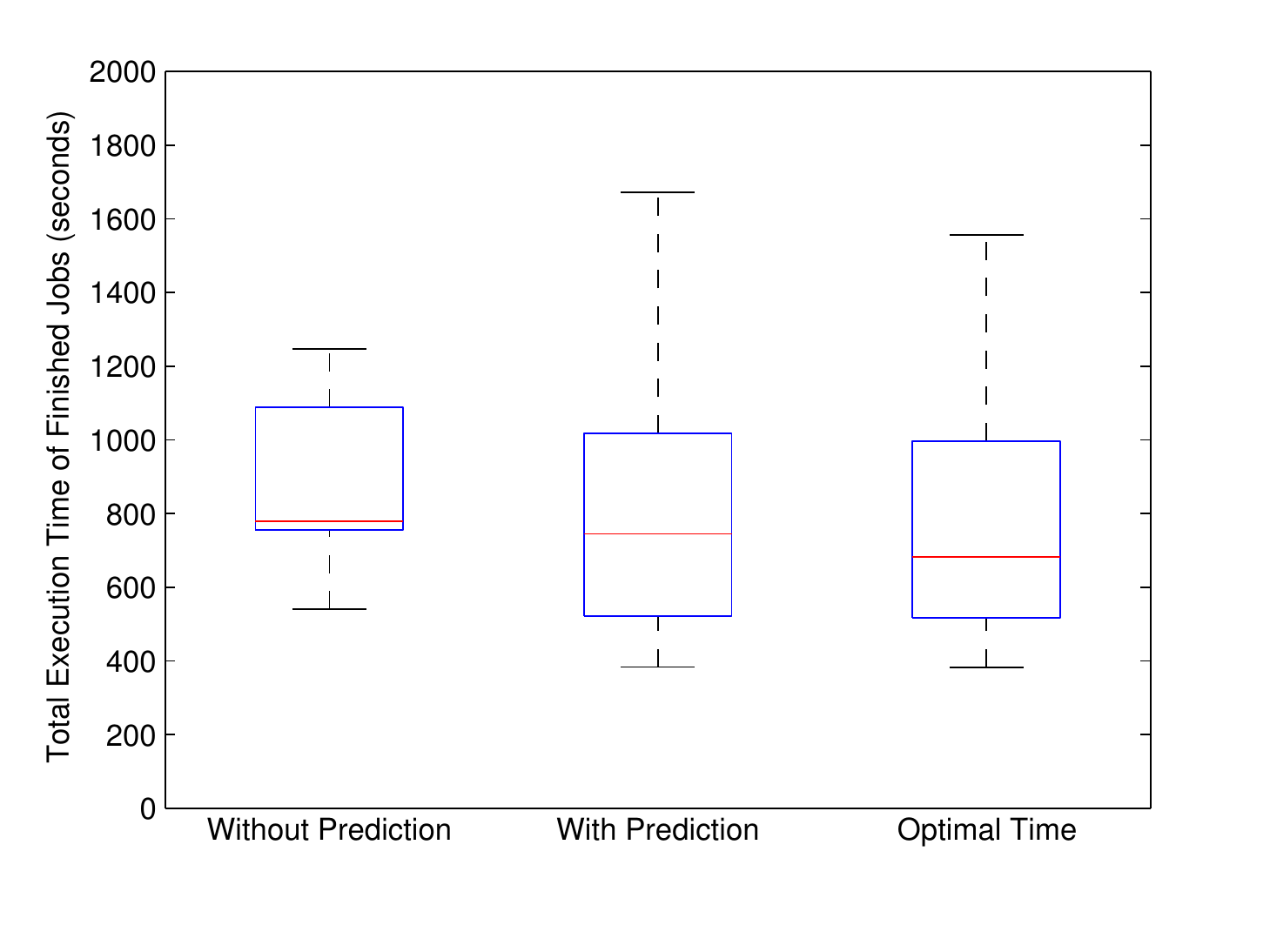}
                \vspace{-7pt}\caption{Batch Jobs}
                \label{fig:batchjobs}
        \end{subfigure}%
        \vspace{0.00mm}
       \begin{subfigure}[b]{0.3\textwidth}
                \includegraphics[width=5cm,height=40mm]{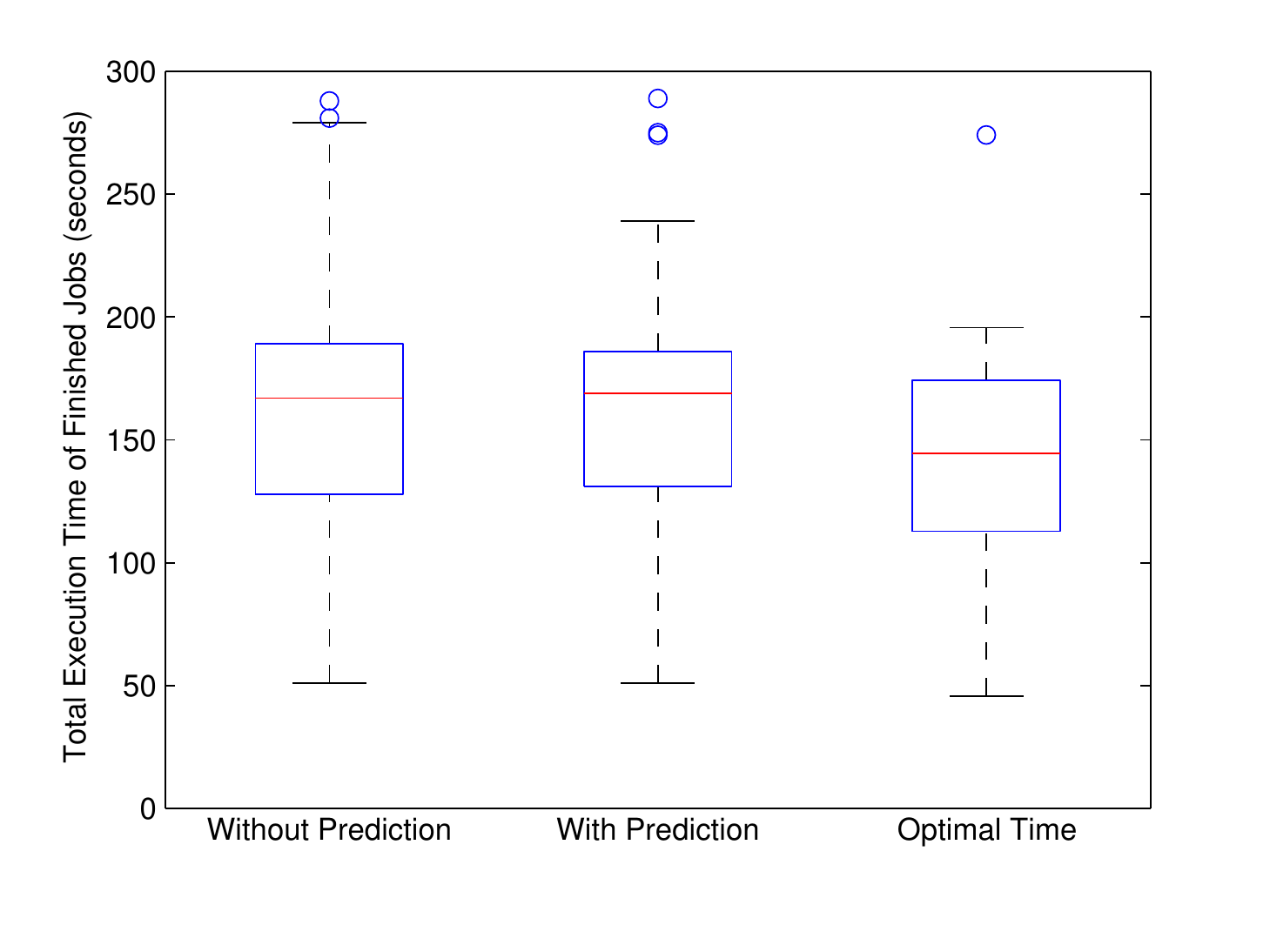}
                \vspace{-7pt}\caption{Mix Jobs}
                \label{fig:mixjobs}
        \end{subfigure}%
        %\vspace{-5pt}
        %\caption{Total Execution Time of Finished Jobs}
        \label{fig:batchtasks}

        \begin{subfigure}[b]{0.3\textwidth}
                \includegraphics[width=5cm,height=40mm]{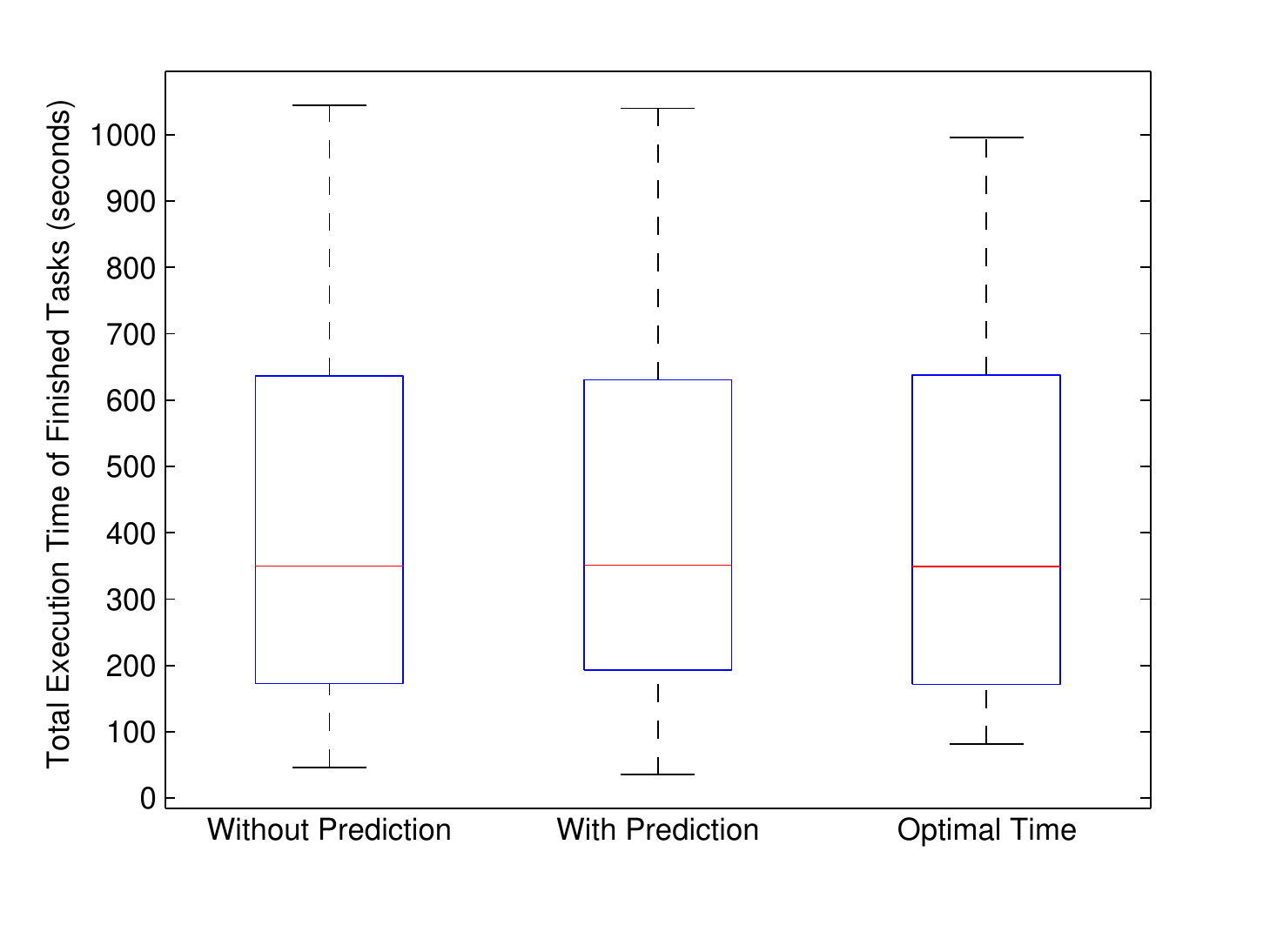}
                \vspace{-7pt}
                \caption{Single Tasks}
                \label{fig:singletasks}
        \end{subfigure}
        \vspace{0.00mm}
       \begin{subfigure}[b]{0.3\textwidth}
                \includegraphics[width=5cm,height=40mm]{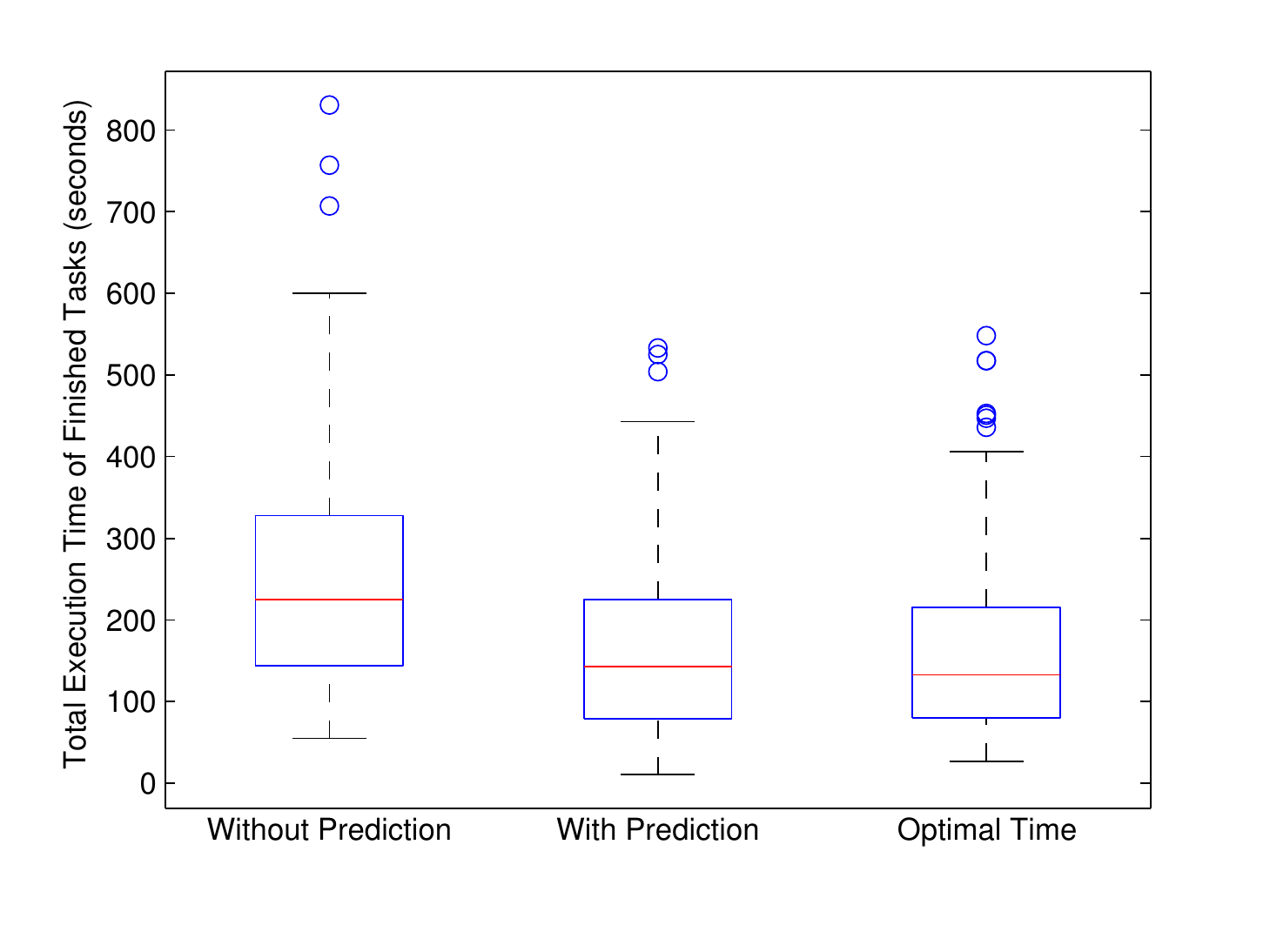}
                \vspace{-7pt}
                \caption{Batch Tasks}
                \label{fig:batchtasks}
        \end{subfigure}%
        \vspace{0.00mm}
       \begin{subfigure}[b]{0.3\textwidth}
                \includegraphics[width=5cm,height=40mm]{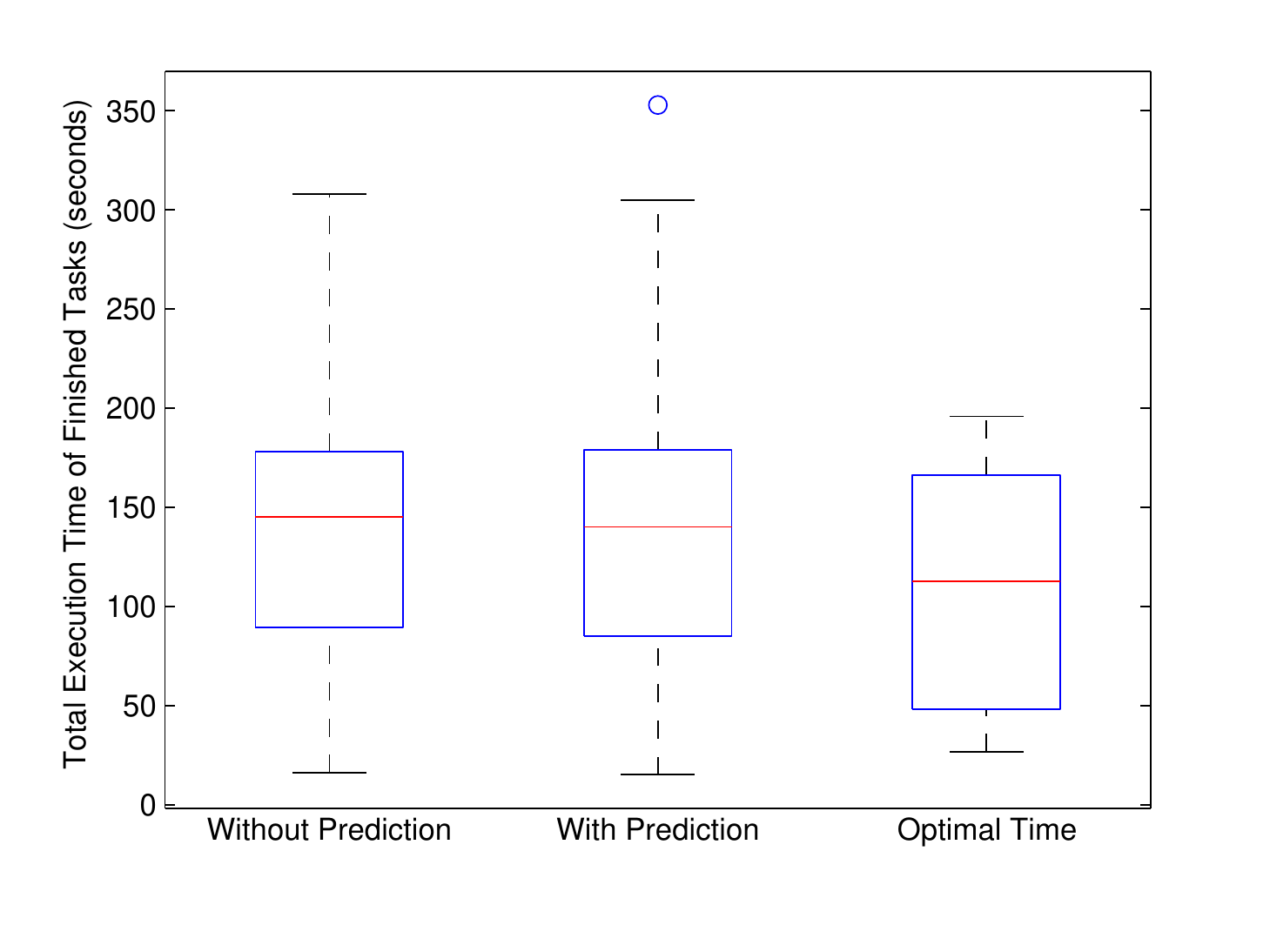}
                \vspace{-7pt}
                \caption{Mix Tasks}
                \label{fig:mixtasks}
        \end{subfigure}%
              \vspace{-5pt}
              \caption{Total Execution Time of Finished  Jobs and Tasks}\label{fig:timetasks}
\end{figure*}

\paragraph*{a) Job Level} \textbf{Overall, the number of finished jobs is increased and the number of failed jobs decreased when extending the GloudSim scheduler with our Random Forest prediction model. In addition, the execution times of the jobs were optimized (the number of rescheduling of failed jobs dropped, reducing the total execution time of the jobs).} % in comparison to the optimal execution time collected using the tool GloudSim.}
The improvement is larger for batch jobs as shown on Figure~\ref{fig:batchjobsfinished} and Figure~\ref{fig:batchjobsfailed} compared to mix jobs as described in Figure~\ref{fig:mixjobsfinished} and Figure~\ref{fig:mixjobsfailed}. For single jobs, the number of finished and failed jobs is almost the same with and without prediction of tasks failure. We explain this result by the fact that our prediction model performs better when a job is composed of multiple dependant tasks; the number of failed dependant tasks and the number of killed dependant tasks are two main characteristics of job failure, as shown on Figure~\ref{Figure:MeanDecreaseGiniJobs}. In general, we conclude that prediction models of tasks can help reduce jobs failure rates because the job scheduling outcome is impacted by the scheduling outcome of its tasks. Moreover, the execution time of jobs was optimized for the batch and mix jobs which can be explained by the reduction of the number of failure events within these jobs, as shown in Figure~\ref{fig:batchjobs} and Figure~\ref{fig:mixjobs}. However, for single jobs, the execution time is the same with or without prediction (see Figure~\ref{fig:singlejobs}). We attribute this outcome to the fact that the distribution of failure events is the same in the two configurations (\ie{} with and without prediction). Overall, reducing the number of failed tasks can help to avoid the starvation problem of tasks waiting on the queue until the successful processing of their dependent tasks, and long scheduling delays in cluster scheduler.

\begin{figure} [ht]
        \centering
        \begin{subfigure}[b]{0.35\linewidth}        %% or \columnwidth
        \centering
        \includegraphics[width=\linewidth]{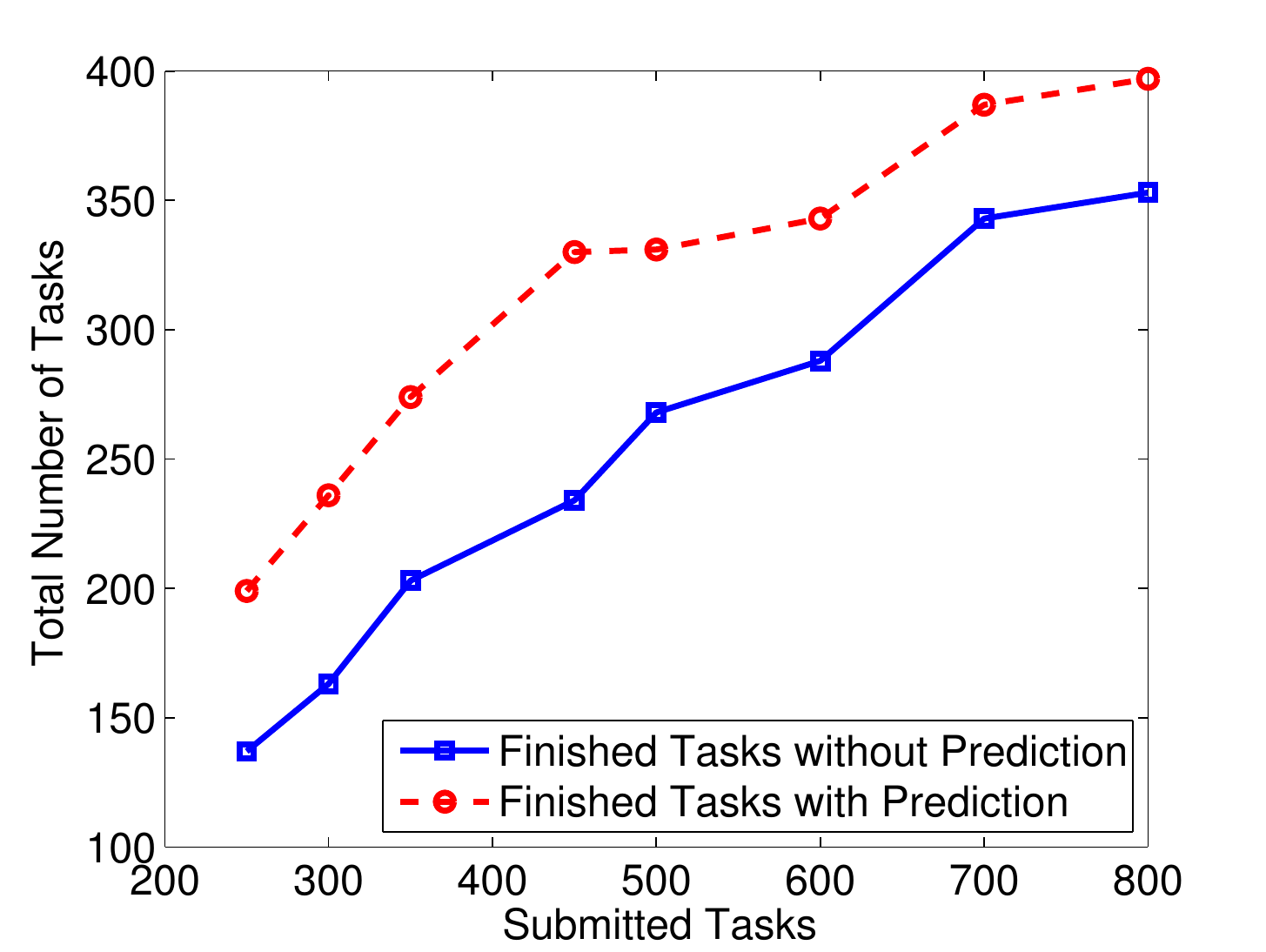}
                \vspace{-5pt}
                \caption{Finished Batch Tasks}
                \label{fig:batchtasksfinished}
        \end{subfigure}
        \begin{subfigure}[b]{0.35\linewidth}        %% or \columnwidth
        \centering
        \includegraphics[width=\linewidth]{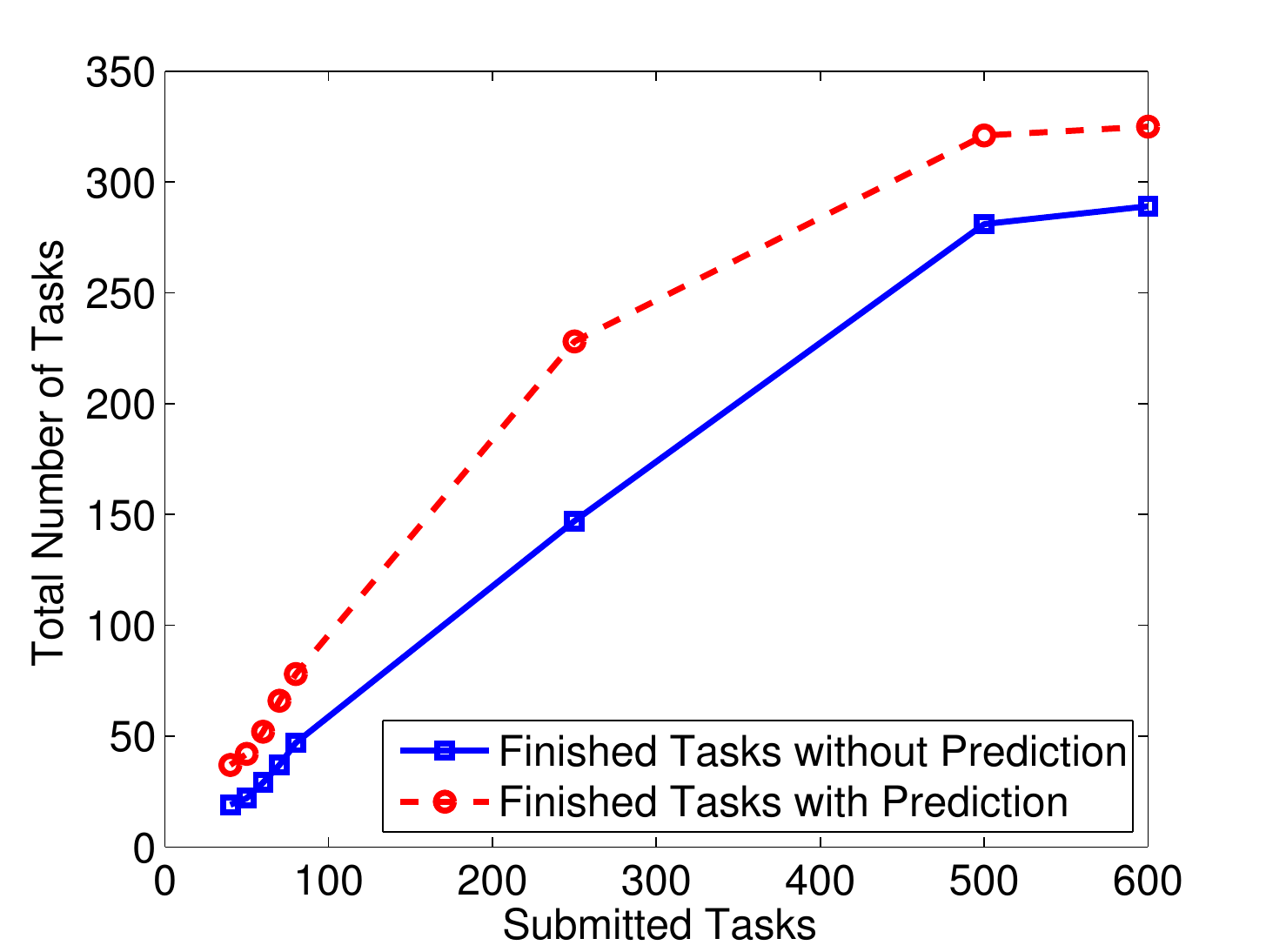}
                \vspace{-5pt}
                \caption{Finished Mix Tasks}
                \label{fig:mixtasksfinished}
        \end{subfigure}%

        \begin{subfigure}[b]{0.35\linewidth}        %% or \columnwidth
        \centering
        \includegraphics[width=\linewidth]{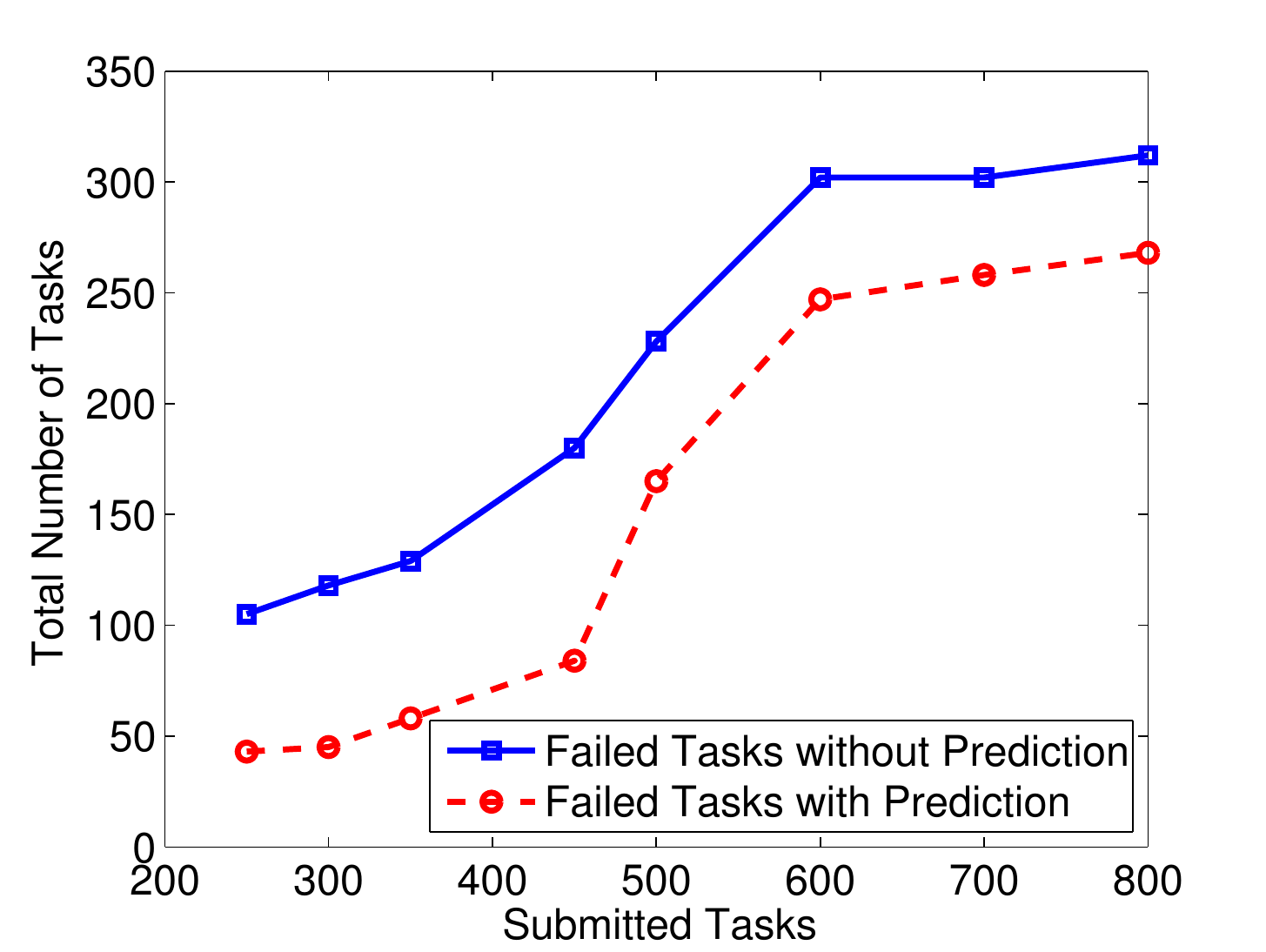}
                \vspace{-5pt}
                \caption{Failed Batch Tasks}
                \label{fig:batchtasksfailed}
        \end{subfigure}
        \begin{subfigure}[b]{0.35\linewidth}        %% or \columnwidth
        \centering
        \includegraphics[width=\linewidth]{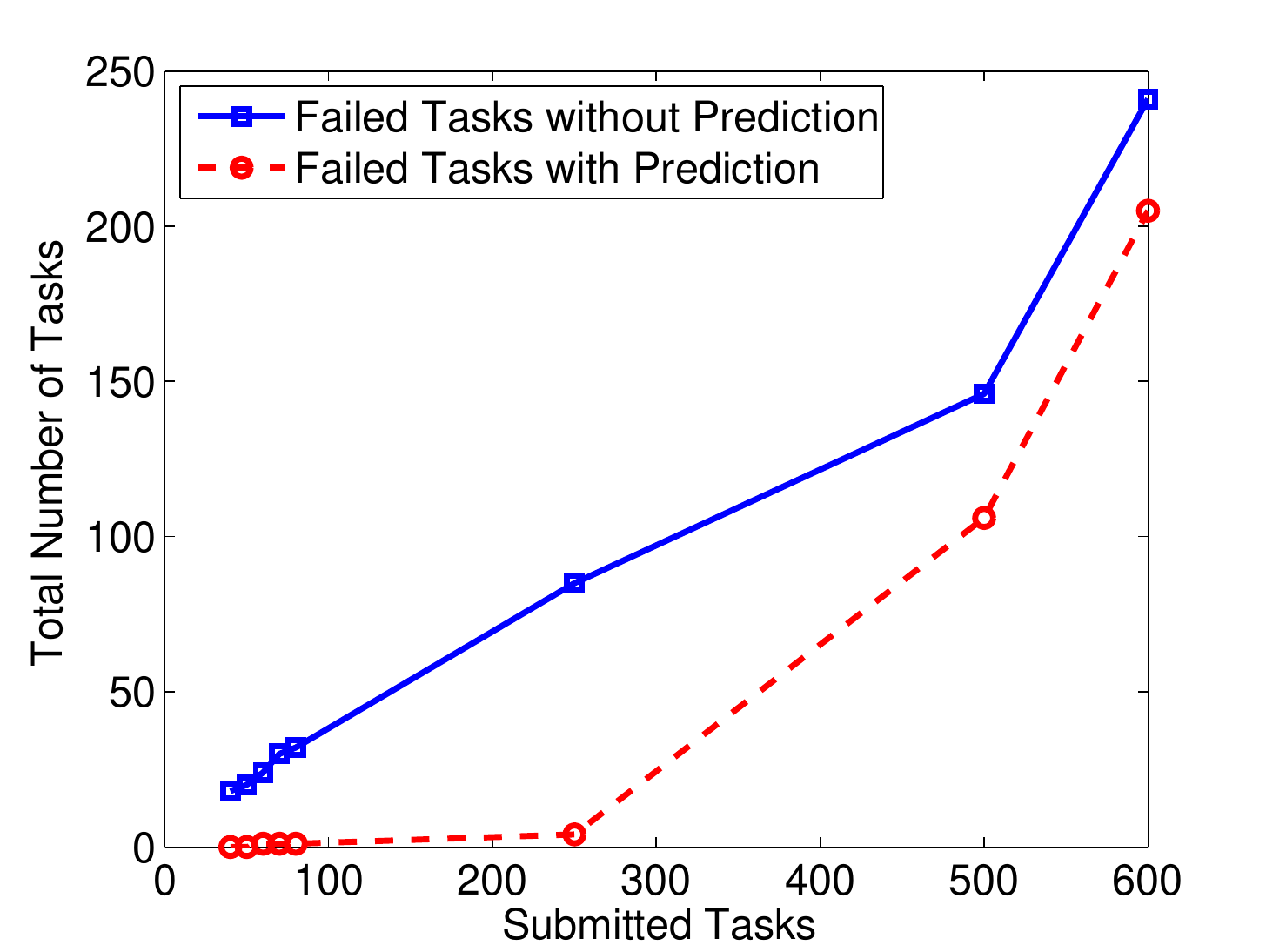}
                \vspace{-5pt}
                \caption{Failed Mix Tasks}
                \label{fig:mixtasksfailed}
        \end{subfigure}%
        \vspace{-5pt}
        \caption{Distribution of Finished and Failed Tasks}
         \label{fig:failedtasks}
\end{figure}

\begin{figure}[ht]
        \centering
       \begin{subfigure}[b]{0.35\linewidth}        %% or \columnwidth
        \centering
        \includegraphics[width=\linewidth]{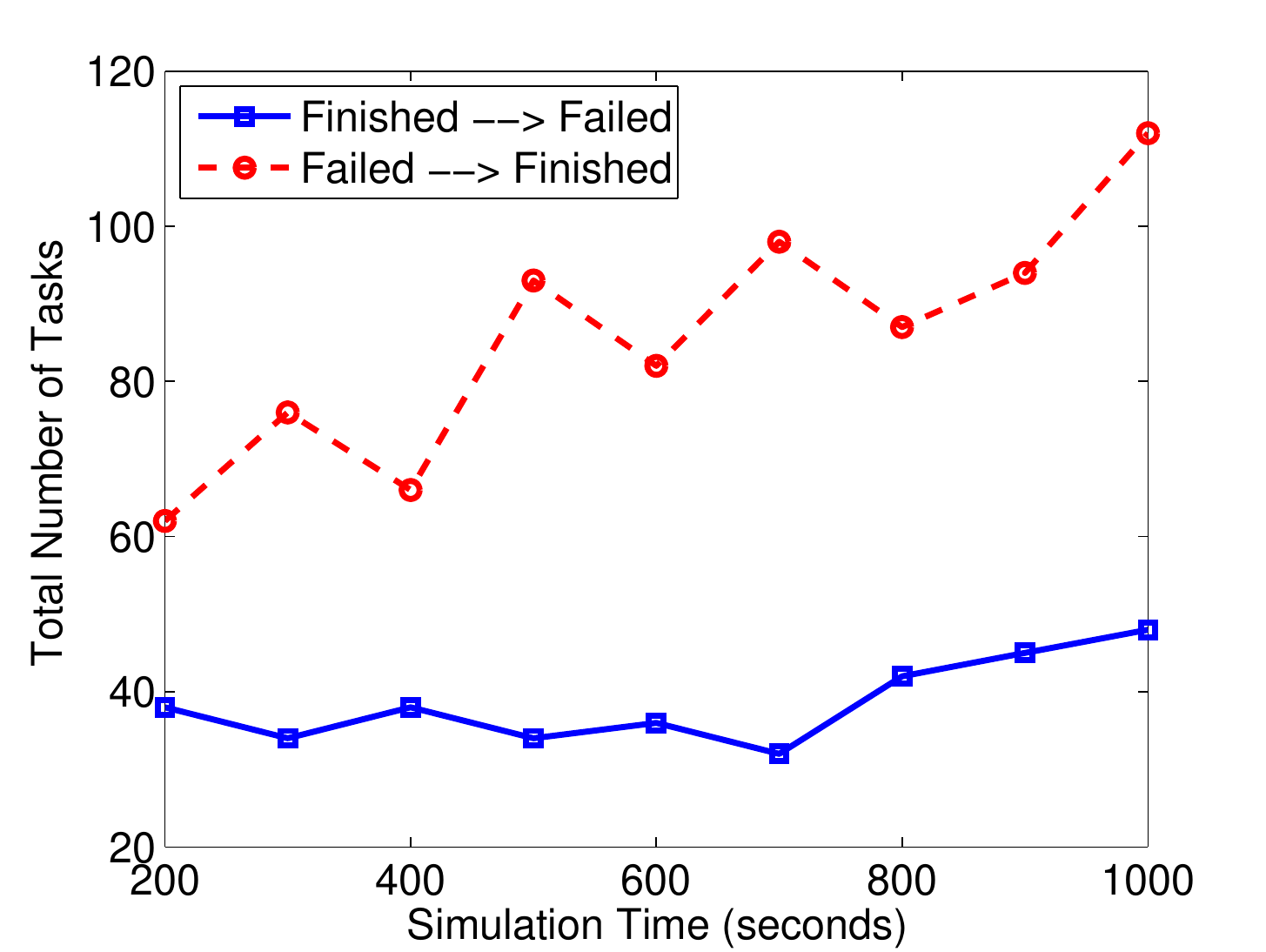}
                \vspace{-5pt}
                \caption{Batch Tasks}
                \label{fig:improvementbatchtasks}
        \end{subfigure}
        \begin{subfigure}[b]{0.35\linewidth}        %% or \columnwidth
        \centering
        \includegraphics[width=\linewidth]{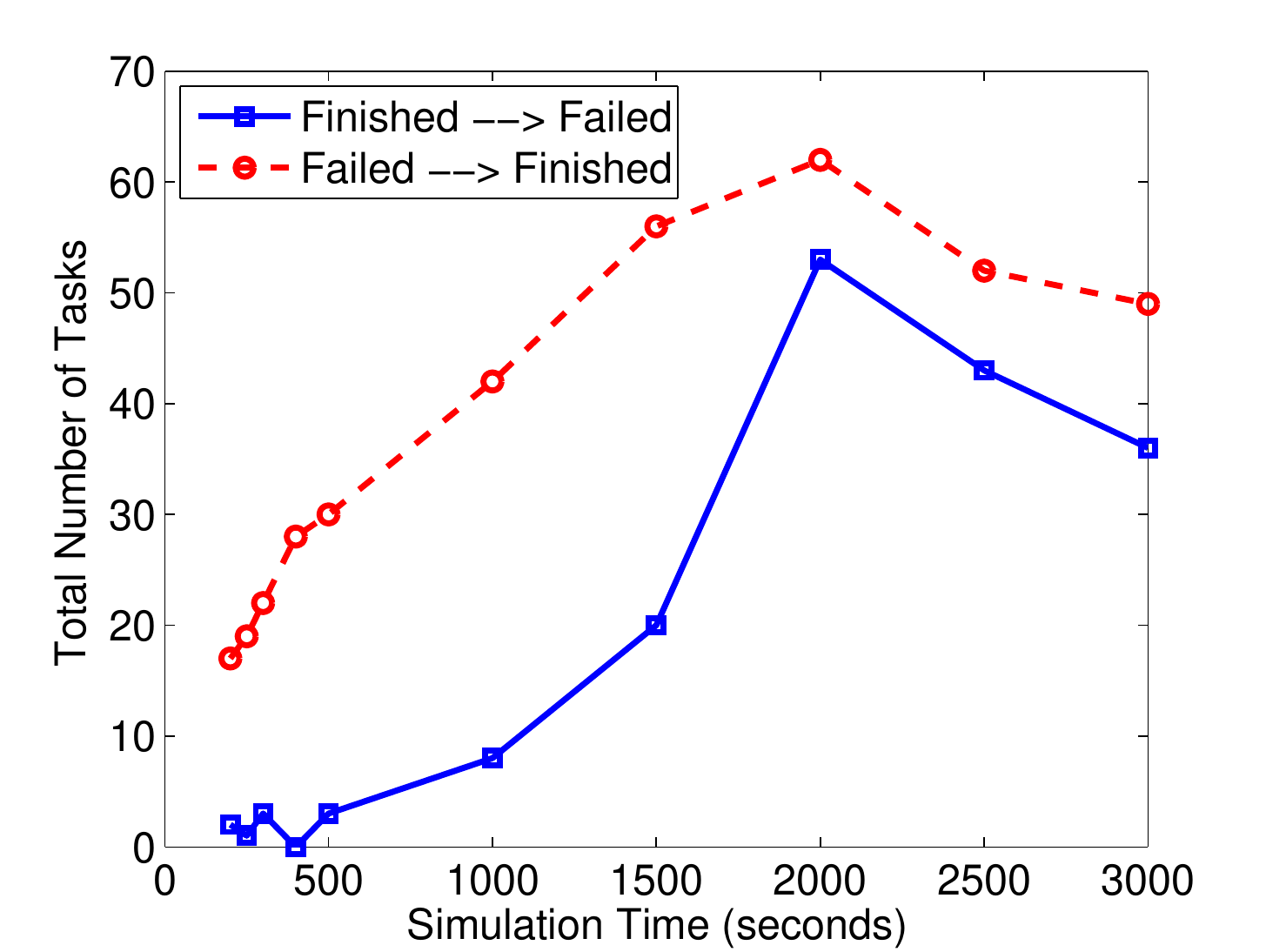}
                \vspace{-5pt}
                \caption{Mix Tasks}
                \label{fig:improvementmixtasks}
        \end{subfigure}%
        \vspace{-5pt}
        \caption{Improvement of the Predictive Model: Task Level}
        \label{fig:finishedjobs}
\end{figure}
\paragraph*{b) Task Level}
%\paragraph{Tasks Level}
\textbf{At task level, we also obtained a reduction of the number of failures and an increase of the number of successful execution when extending the GloudSim scheduler with our Random Forest prediction model.} Similar to jobs, batch tasks show the larger improvements (see Figure~\ref{fig:batchtasksfinished} and Figure~\ref{fig:batchtasksfailed} in comparison to Figure~\ref{fig:mixtasksfinished} and Figure~\ref{fig:mixtasksfailed}). Single tasks show no improvements. %We
Moreover, we noticed that the number of scheduled tasks was improved. This was expected since the prediction model enables the quick rescheduling of tasks that are predicted to fail.
However, we noticed that the number of task failures is still high compared to the number of finished tasks. This is due to the fact that these tasks were failing because of other scheduling constraints (resources, task constraints, etc). Our rescheduling scheme was mainly based on dependencies between tasks but it can be extended to include those other constraints if they are reflected in training data. These failed tasks that we could not predict their failure affect the final scheduling outcome of the jobs.
%potential failed tasks in the queue to wait for some other tasks to finish.
Furthermore, we observed that the execution time was optimized for batch and mix tasks as shown in Figure~\ref{fig:batchtasks} and Figure~\ref{fig:mixtasks} since the submitted tasks were processed and finished without waiting for other submitted or queued tasks to be finished. We explain this improvement by the fact that the scheduler knew in advance which tasks should be scheduled first to ensure the successful processing of the tasks. The execution time of single tasks remained the same as presented Figure~\ref{fig:singletasks}. This is probably due to the fact that our model achieves good prediction, when information about previous killed, evicted and failed dependant tasks are available, which is not the case for single tasks.

% when a job is composed of multiple dependant tasks since the prediction outcomes are based on dependencies between tasks which is not the case of these tasks.

Moreover, we evaluated the improvement of our new scheduler (\ie{} the scheduler extended with our Random Forest model) by computing the number of tasks that were failed without prediction but succeeded (\ie{} their execution finished successfully) when scheduled with the new scheduler.  %task prediction model in terms of number of failed tasks that were finished with prediction.
We also computed the number of tasks that succeeded without prediction but failed when scheduled with the new scheduler.\\ \textbf{Overall, the number of finished tasks was improved by 40\% when scheduling was done with the new scheduler as shown in Figure~\ref{fig:improvementbatchtasks} and Figure~\ref{fig:improvementmixtasks}. 2\% of tasks that succeeded without prediction failed when scheduled with the new scheduler. These failures are due to the false positive predictions of our prediction model. The model is not totally accurate.}

%% file: ApplicationHadoop.tex
\section{Application: Hadoop on Amazon EMR}
\label{sec:application}
\begin{figure*}[th]
\centering
\includegraphics[scale=.4]{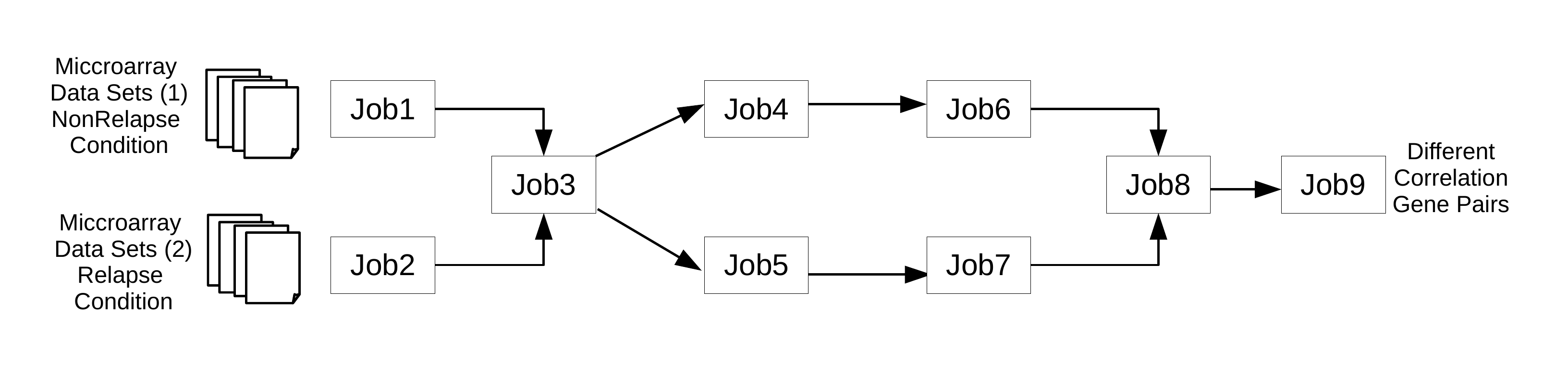}
\caption{Flow of Gene Expression Correlation Analysis Application on Hadoop EMR}
\label{Figure:Applicationflow}
\vspace{-5pt}
\end{figure*}
%Describe the approach on Hadoop + Application
%Following our methodology described in Section \ref{methdology}, 
We implemented and deployed the prediction models from Section \ref{methdology} on Amazon EC2, extending the standard scheduler of the Hadoop framework on Amazon EMR. To evaluate the performance of these extended schedulers,
%
%We applied our proposed methodology described in\ref{methdology} on Hadoop by running some jobs on its nodes and then get the relevant information formthe log files and build the predictive model which determines whether a job will fail or not. After, we added this predictve schemeto Hadoop scheduler and evaluate its performance compared it default implementation.
%To do so, 
we selected an application used for gene expression correlations analysis in the context of Breast cancer research. The application is used to uncover factors causing breast cancer by identifying differential gene expressions between different conditions (\eg{} cancerous versus normal cells).
This application was recently parallelized by Tzu-Hao \emph{et al} \cite{[Chang-BreastCancer]}, using the MapReduce programming model and deployed on Hadoop \cite{[EMR]}. %deployed on  in breast cancer research
%
%an application used to perform gene expressions analysis for breast cancer disease \cite{[Chang-BreastCancer]}. To decrease the computation time of this application, Tzu-Hao et al. implemented MapReduce program to perform this application and they deployed on Hadoop\cite{[EMR]}. 
%The application is used to discover the factors causing this disease by identifying the differential gene expressions between conditions (e.g., cancer versus normal cells). Results' accuracy of this application is crucial in this domain and 
Since the application performs sensitive analysis, any job failure on Hadoop may lead to inaccurate information and wrong conclusions about the disease.
%As described in \cite{[Chang-BreastCancer]}, 
The application is composed of 9 jobs \cite{[Chang-BreastCancer]} shown in Figure~\ref{Figure:Applicationflow}. These jobs are dependent and any failure occurrence may propagate inaccurate results and incur delays that will affect the total execution time of the application.
%Describe the Environment on AmazonEMR:
We emulated the behaviour of this application by running the same flow (presented on Figure~\ref{Figure:Applicationflow}) using the \emph{wordcount} example provided by Apache with Hadoop as job unit; linking the output of these dependent jobs together as on Figure~\ref{Figure:Applicationflow}, to obtain the final output.

We ran the analysis on Amazon EMR using 4 machine instance of type \textit{m3.large( ECPU=6.5, VCPU=2, MEM=7.5 GB, Instance Storage=32, Network Performance= Moderate)}. The first machine was the master node submitting the jobs to two other machines considered as the workers. The last machine was the secondary master node.
We ran different simulations on these machine instances to collect log files that we parsed to extract attributes described in Section \ref{methdology} and train the prediction models. %build the predict the potential scheduling  outcome of job. After removing the correlated attributes, we found that the scheduling outcome is highly dependent on the number of failed job in the past of job. So the more the job is characterized by more dependent failed jobs in the past the more it has the probability to be failed. Thus, it is fundamental to early predict these failure and reschedule the potential failed jobs in order to not propagate the failure and extra delays.
%
%describe the Learning Part
We compared the performance of the six models described in Section \ref{methdology} and found that the best results are achieved with Neural Networks, \ie{} accuracy (72.8\%), precision (97.2\%) and recall (72.7\%). We used different training and testing data sets when assessing the performance of the models. Table~\ref{Table:ResultsHadoop:R} summarizes the results achieved by the six models. %  achieves the best result in this case, obtained the bestThen, we used different training and testing dataset to perform 10-cross validation in order to determine the predictive algorithm having the highest accuracy, recall and precision. The obtained results are described in TAble~\ref{Table:ResultsHadoop:R}.  We found that Neural Network is performing good and it gives good results in terms of accuracy (72.8), precision (97.2) and recall (72.7). Therefore, we will use to Neural Network algorithm to implement predictive scheme of job scheduling failure on Hadoop.
%This scheme enables the scheduler to early identify if a job will fail or not based on the collected data from the cloud environment. If the submitted job will fail, then the scheduler should select the appropriate rule to reschedule based on its dependencies with other jobs.

\begin{table}[ht]
\caption{Accuracy, Precision, Recall (In \%) obtained from different Algorithms: (Random K-Cross Validation K=10)}
\centering \scriptsize
\label{Table:ResultsHadoop:R}
\begin{tabular}{|> {\centering\arraybackslash}p{3.5cm} |>{\centering\arraybackslash}p{2cm} |>{\centering\arraybackslash}p{2cm}|>{\centering\arraybackslash}p{2cm}|}
\hline
 \textbf{Algo. }&  \textbf{Acc.}  & \textbf{Pre.} & \textbf{Rec.} \\ [0.5ex] % inserts table
\hline\hline
  Tree &  41.9  & 84.1 & 42.7 \\\hline
   Boost & 38.1  & 72.8 & 39.2 \\\hline
   Glm  & 26.5  & 77.8 & 26.0 \\\hline
   CTree  & 61.8  & 89.7 & 54.9 \\\hline
   Random Forest & 24.0  & 69.9 & 23.7 \\\hline
   Neural Network  & 72.8 &  97.2 & 72.7 \\\hline
\end{tabular}
\end{table}

Using Hadoop's scheduler extended with the Neural Network prediction model, refreshing its scheduling policies every 5 minutes, we performed different simulations by submitting jobs to worker nodes and injecting an early failure on Job1, a late failure on Job8 and 2 mixed failures (late and early) on Job1 and Job8. For each of these simulations, we measured the total execution time and the total number of failed jobs and compared the obtained results with those of the default Hadoop scheduler.
%
%In order, to demonstrate the practicality of  of our the selected predictive model, we performed different simulation on which we submitted different jobs to the worker nodes then inject different type of failure by killing some jobs and evaluate the performance of the running jobs. We injected the early failure on Job1, the late failure on Job8 and 2 mixed failure (late and early) on Job1 and Job8. Then, we measured the total execution time of the default Hadoop scheduler and the three scenarios of failure.
We also measured the execution time of the jobs for different numbers of learning iterations.

\begin{figure}[th]
\centering
\includegraphics[scale=.5]{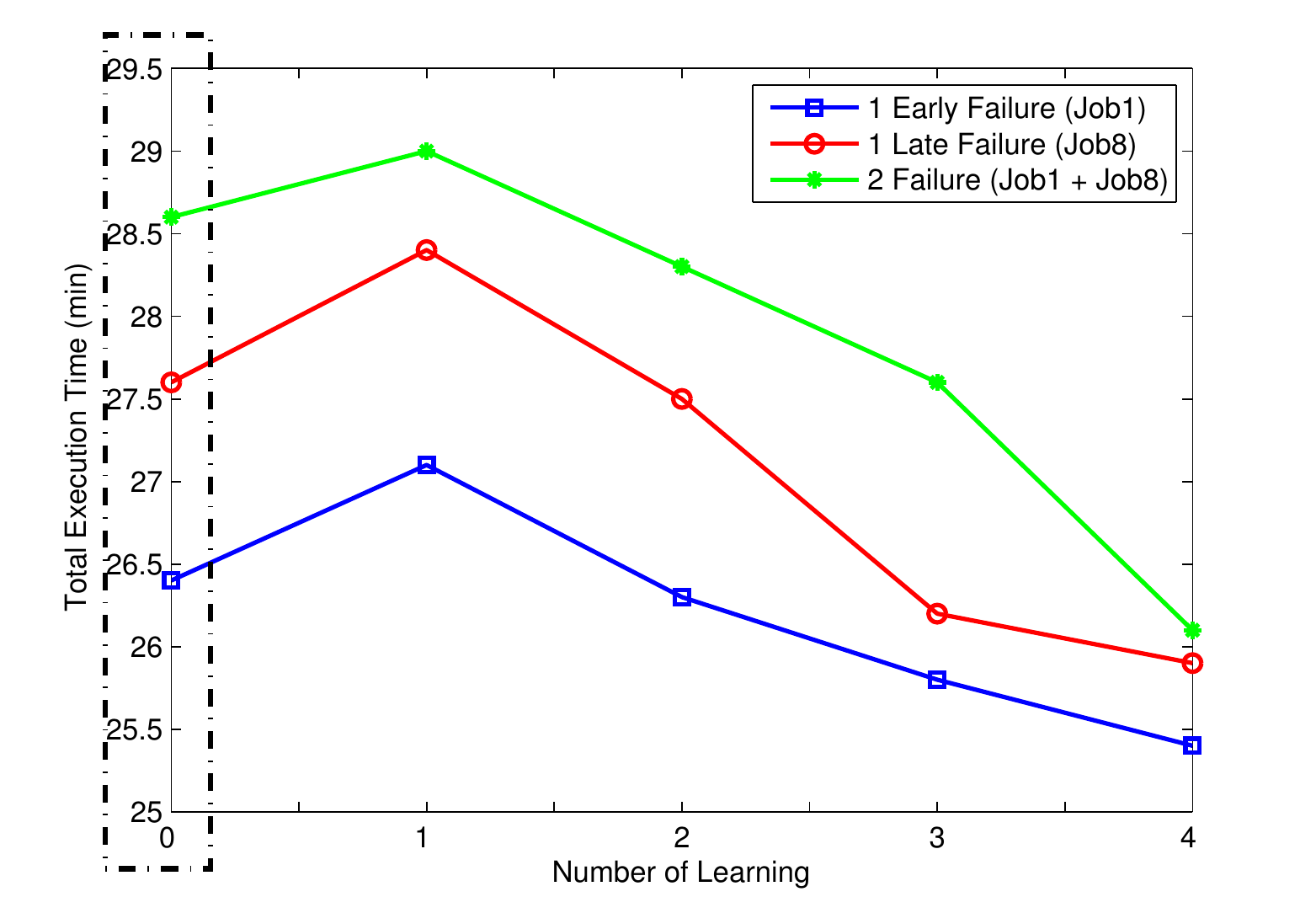}
\caption{Total Execution Time of Hadoop Jobs}
\label{Figure:ExecutionTimeJobsHadopp}
\vspace{-5pt}
\end{figure}

\begin{figure*}[ht]
        \centering
        \begin{subfigure}[b]{0.3\textwidth}
                \includegraphics[width=5cm,height=40mm]{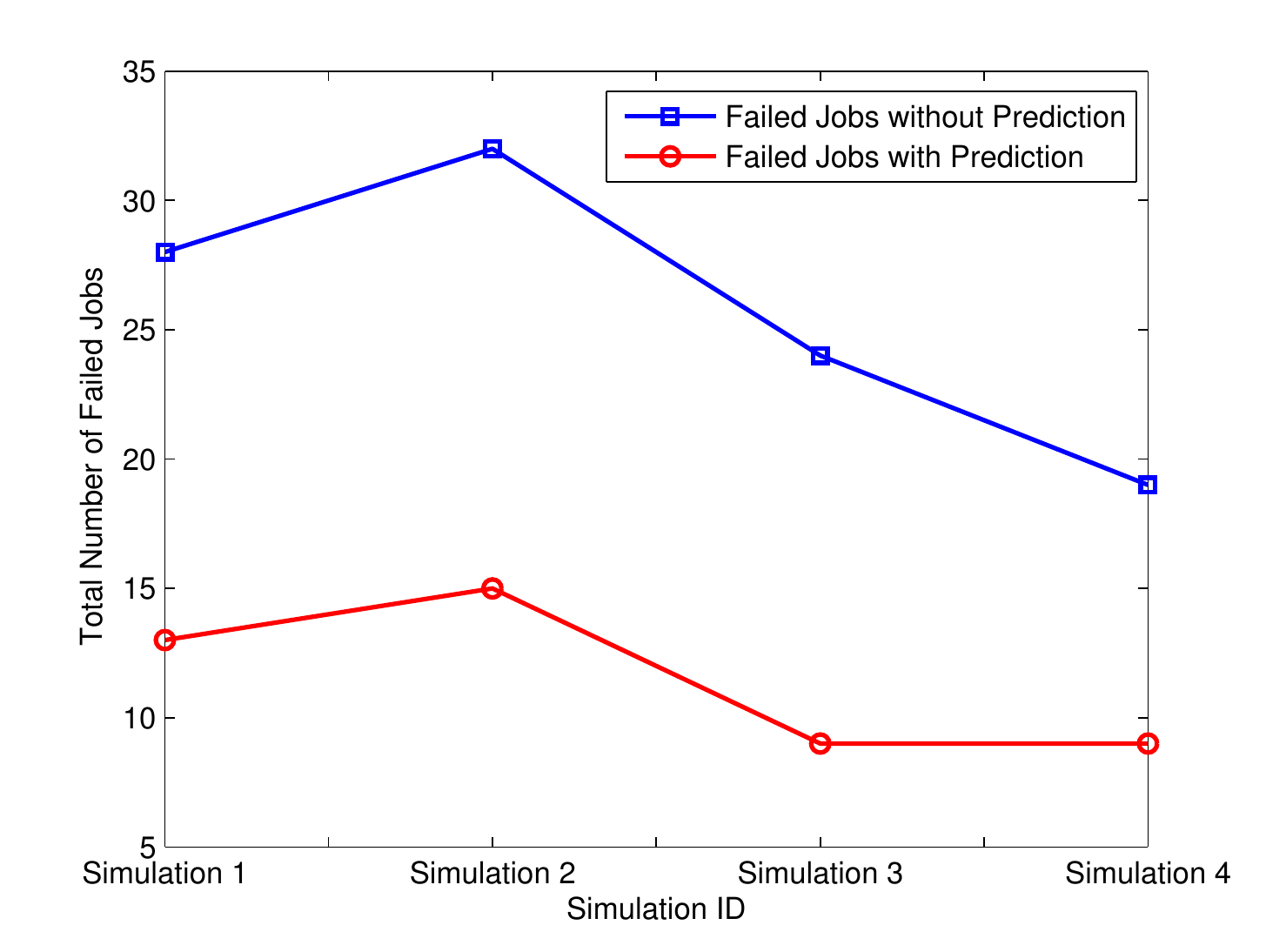}
                \vspace{-7pt}\caption{Early Injected Failure}
                \label{fig:failure1}
        \end{subfigure}
        \vspace{0.00mm}
       \begin{subfigure}[b]{0.3\textwidth}
                \includegraphics[width=5cm,height=40mm]{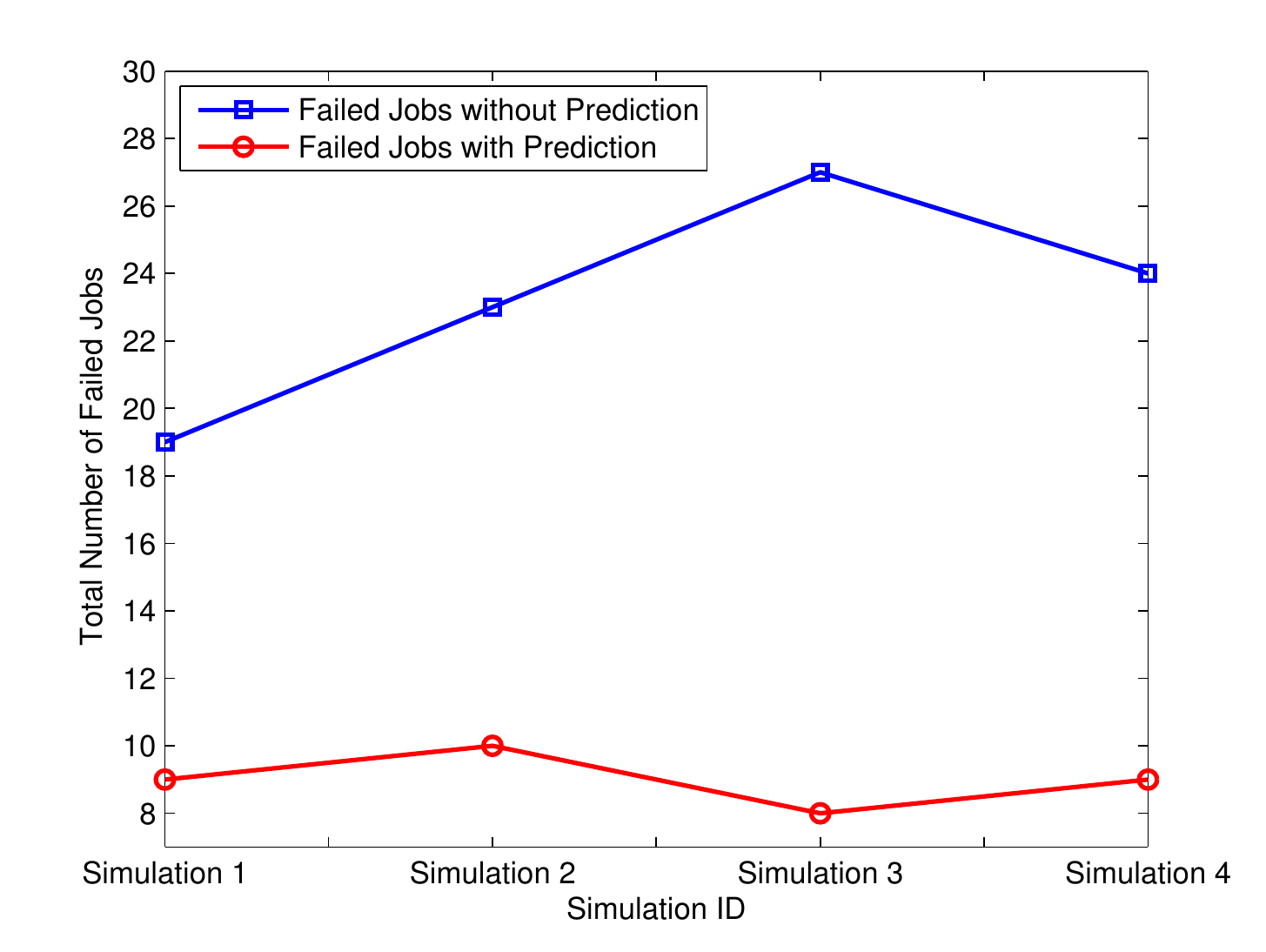}
                \vspace{-7pt}\caption{Late Injected Failure}
                \label{fig:failure2}
        \end{subfigure}%
        \vspace{0.00mm}
       \begin{subfigure}[b]{0.3\textwidth}
                \includegraphics[width=5cm,height=40mm]{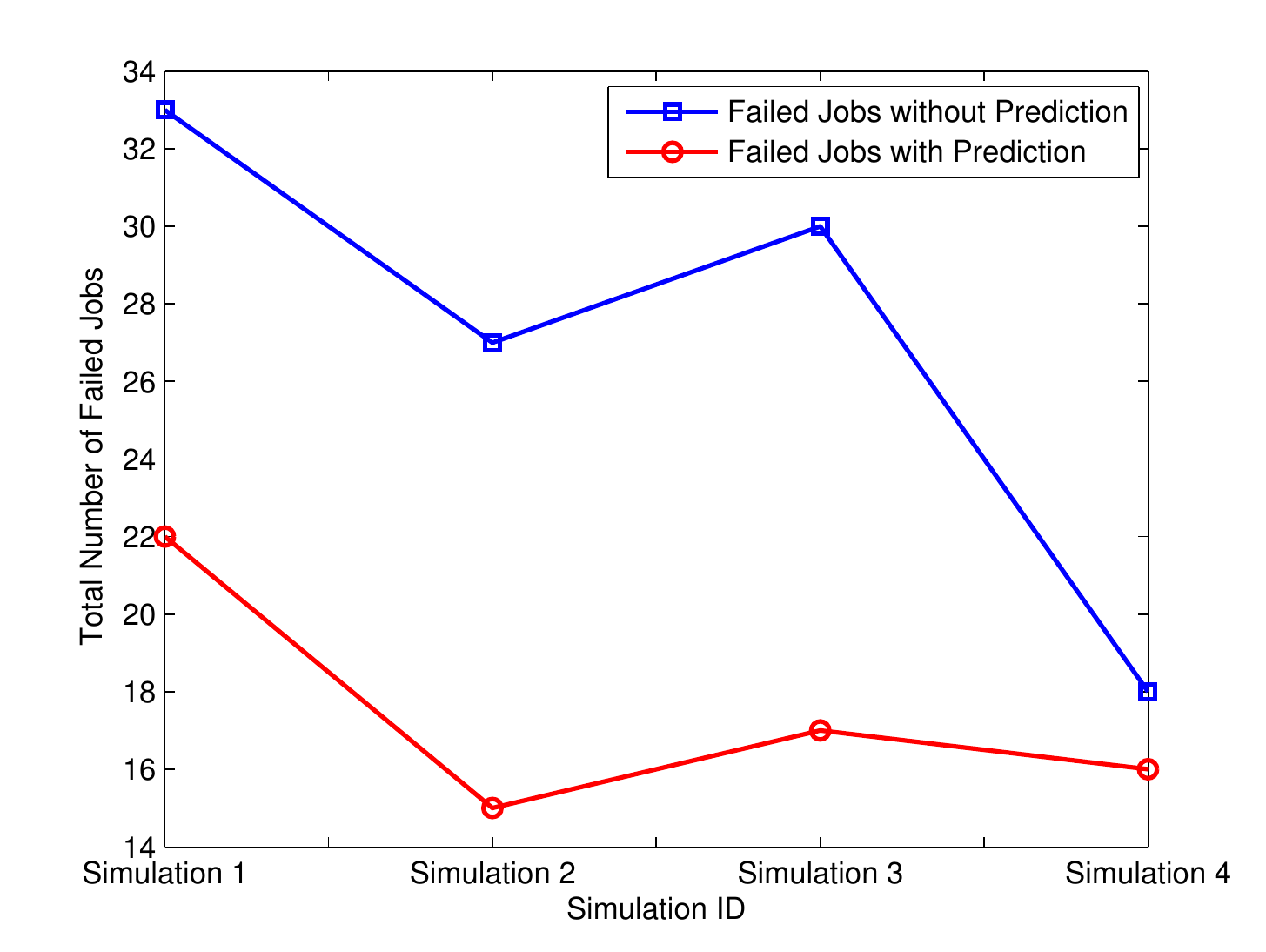}
                \vspace{-7pt}\caption{2 Injected Failure}
                \label{fig:failure3}
        \end{subfigure}%
         \vspace{-5pt}
         \caption{Total Number of Failed Jobs in Hadoop}
         \label{fig:FailedJobsHadoop}
\end{figure*}

The Hadoop's scheduler extended with the Neural Network prediction model can reduce the number of job failures by up to 45\% (see Figure~\ref{fig:failure1}, Figure~\ref{fig:failure2} and Figure~\ref{fig:failure3}). In our proposed scheduling scheme, if the job satisfies its dependency requirement, it will be submitted to the worker to be executed or it will be rescheduled and restarted from the beginning to ensure its successful completion. However, we noticed that there are still some jobs which are still failing although they were predicted to be completely finished. This is due to the failure related to cluster environment (errors occur while executing the job, insufficient resources, long running tasks, etc). Moreover, the execution time of jobs is significantly reduced with more learning (Figure~\ref{Figure:ExecutionTimeJobsHadopp}). We explain this results by the fact that more learnings (\ie{}., more trainings of the model on larger historical data) improve the performance of the prediction model, enabling it to reschedule earlier multiple jobs that would have failed.

%% file: threats.tex
\section{Threats to Validity}
\label{sec:threats}
This section discusses the threats to validity of our study following the guidelines for case study research~\cite{robert2002case}.

\emph{\textbf{Construct validity threats}} concern the relation between theory and observation. Our modelling approach assumes that tasks and jobs characteristics alone can explain scheduling outcomes, when in reality, this may not be the case. It is possible that other factors such as scheduling class or resources allocation strategy also play a role in scheduling decisions. However, in our data set we found a low correlation between scheduling class and scheduling outcomes. According to ~\cite{[8]}, this low correlation is due to the fact that the scheduling class which represents the latency-sensitivity of a task/job mostly affects local machine policies for accessing cluster resources. They are therefore more likely to affect resource usages than scheduling outcomes. Also, in our data set, assigned resources were too high compared to the requested resources (this is explained by the overbooking strategy adopted by Google ~\cite{[16]}) nullifying the impact of resource allocation on scheduling outcomes. Also, our modelling did not considered task constraints, which specify the machines on which a task can run. In future work, we plan to examine the relation between task constraints and scheduling outcomes.
%In future work we plan to investigate.....\Foutse{complete?}.
%
Another construct validly threat concerns the size of the data set on which our results are obtained (we randomly sampled 2\% of tasks and jobs contained in the Google data set). However, since we obtained consistent results for different samples of the same Google data set, we are confident that our findings can hold on the whole Google data set. %ur data sampling.
We analyzed only 2\% of tasks files contained in the Google dataset. However, to ensure that our findings hold for the remaining dataset, we collect 10\% additional sample from the same dataset and obtained the same results.
%have verify our results on multiple samples resampled

\textbf{\emph{Internal validity threats}} concern our selection of subject systems, tools, and analysis method. Although the Google data set used in this study may not contain all the different kinds of task and jobs used in the industry, it represents the execution of real applications from a major company (\ie{} Google). The GloudSim scheduler used in \textbf{RQ3} does not represent of all existing schedulers in the industry. However, it is designed to reproduce the scheduler used in some Google clusters. Moreover, the tool kit GloudSim has already been used successfully in many research projects ~\cite{[25]} ~\cite{[26]}. When running the gene expression correlations application on Hadoop in Section \ref{sec:application} we used wordcount data for the jobs instead of the gene data and we injected failures in the nodes. Although this may not represent the natural execution of that application, we believe that it enables a good assessment of the performance of our extended schedulers in reducing job failures. %

\textbf{\emph{Conclusion validity threats}} concern the relation between the treatment and the outcome. We paid attention not to violate assumptions of the constructed statistical models.

\textbf{\emph{Reliability validity threats}} concern the possibility of replicating this study. Every result obtained through empirical studies is threatened by potential bias from data sets \cite{menzies2007data}. We provide all necessary details needed to replicate our study. The Google data set is publicly available for study.

\textbf{\emph{External validity threats}} concern the possibility to generalize our results. Our study is based on large-scale data (\ie{} \textit{158 GB}) collected from Google clusters. Nevertheless, further validation on larger and diverse sets of tasks and jobs is desirable.

%% file: relatedwork.tex
\section{Related Work}
\label{sec:relatedwork}
There is a large body of research  that aimed to characterize the task and jobs contained in the Google cluster traces used in this paper. We classify these works into the three following categories.
\subsection{Workload Characterization}
Di et al.~\cite{[3]}, studied the resources utilisation of applications from the Google data set, using the \textit{K}-means clustering algorithm. They studied whether the resources within the cluster can execute the batch tasks or not.
Liu et al. \cite{[4]} used the traces files to study the main characteristics of the machines used to perform the tasks and jobs. They also analysed the impact of machine workload on the overall resources utilization to characterize the machines management system of the cluster. Chen et al. \cite{[6]} proposed an approach to classify workloads based on cloud performance using a tool named Statistical Workload Analysis and Replay for MapReduce (SWARM) to evaluate the impact of batched and multi-tenant execution on jobs latencies and the cluster utilization. Kavulya et al.~\cite{[7]} analysed job processing in Hadoop and proposed an analytical model to predict the total completion time of a job. %There are other works that used traces collected from Yahoo! and M45 production clusters to analyse the performance of clusters. 
Although, workload characterization can improve clusters' management, it is also important to characterize jobs and tasks schedulings, for example by analyzing the relations between workloads and scheduling outcomes.

\subsection{Scheduling Characterization}
The characterization of scheduling events has been the focus of many
workload analysis studies. Recently, \cite{[12]} addressed the batch jobs scheduling in distributed data centres and proposed GreFar to optimize the energy cost and fairness across different clusters which are characterized by scheduling delays constraints. Zhang et al. used the Google cluster data to propose Harmony, a heterogeneity-aware framework that can minimize scheduling delays and the total energy consumption by controlling the number of machines that are provisioned~\cite{[11]}\cite{[14]}. The performance of Harmony was found to be better than GreFar \cite{[12]}.
In \cite{[10]}, Sharma et al. showed that task placement have a large impact on scheduling delays: task waiting time can be increased by a factor of 2 to 6 due to the cluster and task constraints. They proposed a methodology that takes into account resources requirement and task placement.
\subsection{Failure Analysis and Prediction}
Failure analysis and prediction have become popular in researches on distributed systems, since they allow for early identification of failure and can improve the performance of the cluster. Fadishei et al.~\cite{[18]} used the Grid Workload Archive project to analyse the correlation between job failures and resources attributes (e.g., resources utilisation and scheduler characteristics). They found that scheduler load, execution hour of day and CPU-intensity are among the most factors that can affect failure rates.
Ganesha was proposed by Pan et al~\cite{[19]} as a black-box tool to identify failures between faulty and normal nodes in MapReduce. Xin Chen et al. used Google trace files to identify and predict jobs failure in batch applications. They used Recurrent Neural Networks to perform their predictions. This model was able to reduce resources utilisation by between 6\% and 10\%~\cite{[21]}\cite{[22]}. They recommend that predicted failed tasks be killed immediately without processing, in order to avoid wasting resources.
However, killing predicted failed tasks is likely to affect the overall performance of a cloud application. A better decision would be to reschedule the tasks quickly on appropriate clusters with adequate resources.
To the best of our knowledge, our work is the first that proposes an approach to predict and reschedule failed tasks in order to improve the performance of cloud systems.
In addition, our work evaluates the performance of many statistical models in predicting failed tasks. Also, we show that Random Forest achieves better results compared to Neural Networks, both in terms of precision and recall.

%% file: conclusion.tex
\section{Conclusion}\label{sec:conclusion}
Task scheduling is an important issue that greatly impacts the performance of cloud computing systems. In this paper, we examined task failures in Google clusters data and found that 42\% of the jobs and 40\% of the tasks were not finished successfully. We noticed that a job often fails because of the failures of some of its tasks, and tasks also fail because of the failure of dependent tasks. We investigated the possibility of predicting the scheduling outcome of a task using statistical models and historical information about the execution of previously scheduled tasks and found that Random Forest models can achieve a precision up to 97.4\%, and a recall up to 96.2\%. We also extended the schedulers implemented in GloudSim and Hadoop to incorporate task failure predictions; the goal being to achieve early rescheduling of potential failed tasks (\ie{} early on before their actual failing time). We compared the scheduling performance of the new scheduler and the original scheduler implemented in GloudSim, in terms of execution times and numbers of finished tasks and jobs, and found that the number of finished tasks (respectively jobs) can be increased by up to 40 \% (respectively 20 \%) and the execution time reduced by the new scheduler. In the case of Hadoop, the new scheduler can reduce the number of job failures by up to 70\% with an overhead time of less than 5 minutes. Cloud service providers could improve the performance of their task scheduling algorithms by extending them with our proposed failure prediction models. Since the extra layer of prediction can have an impact on the performance of cloud applications (\ie{} training and applying the proposed prediction model can cause delays in scheduling decisions), although we found it to be less than 5 minutes in our case study on Hadoop, in future work, we plan to examine in details the %In this study we noticed that
trade-offs between precision and execution time when selecting a prediction model, as well as the frequency at which the predictions should be performed in order to ensure optimal scheduling response times.